\documentclass[conference,compsoc]{IEEEtran}
\IEEEoverridecommandlockouts
\usepackage{cite}
\usepackage{eso-pic} 
\usepackage{amsmath,amssymb,amsfonts}
\usepackage{graphicx}
\usepackage{microtype}
\usepackage{textcomp}
\usepackage{url}
\usepackage{booktabs}
\usepackage{multirow}
\usepackage[table]{xcolor}
\usepackage{xurl}
\usepackage{hyperref}
\usepackage{seqsplit}
\usepackage{tikz}
\usepackage{pgfplots}
\usepackage[most]{tcolorbox}
\usepackage{tabularx}
\usepackage{float} 
\pgfplotsset{compat=1.18}
\usetikzlibrary{positioning,arrows.meta,shapes.geometric}

\usepackage{caption}

\begin{document}

\AddToShipoutPictureFG*{%
  \AtPageUpperLeft{%
    \put(54,-27){%
      \raisebox{-\height}[0pt][0pt]{%
        \begingroup
        \setlength{\fboxsep}{5pt}%
        \setlength{\fboxrule}{0.45pt}%
        \fcolorbox{black!50}{white}{%
          \parbox{\dimexpr504pt-2\fboxsep-2\fboxrule\relax}{%
            \footnotesize\raggedright
            This paper has been accepted for publication at ACSAC 2026.
            If you wish to cite this work, please use the following reference:
            Hiroki Nakano, Takashi Koide, and Daiki Chiba,
            ``TIBlender: Early-Warning Threat Intelligence from
            Cross-Platform Social Media Evidence,''
            \emph{Proceedings of the 42nd Annual Computer Security
            Applications Conference (ACSAC 2026)}, 2026.
          }%
        }%
        \endgroup
      }%
    }%
  }%
}

\title{TIBlender: Early-Warning Threat Intelligence from \\Cross-Platform Social Media Evidence}

\author{
    \IEEEauthorblockN{Hiroki Nakano}
    \IEEEauthorblockA{
    \textit{NTT Security Holdings Corporation}\\
    \textit{\& NTT, Inc.}\\
    Tokyo, Japan\\
    hi.nakano.sec@gmail.com
    }
    \and
    \IEEEauthorblockN{Takashi Koide}
    \IEEEauthorblockA{
    \textit{NTT Security Holdings Corporation}\\
    \textit{\& NTT, Inc.}\\
    Tokyo, Japan\\
    takashi.koide@security.ntt
    }
    \and
    \IEEEauthorblockN{Daiki Chiba}
    \IEEEauthorblockA{
    \textit{Tokyo Metropolitan University}\\
    Tokyo, Japan\\
    daiki.chiba@ieee.org
    }
}

\maketitle
\thispagestyle{plain}
\pagestyle{plain}

\begin{abstract}
Cyber threat signals are fragmented across multiple social media platforms, yet no existing approach has fully automated their integration into actionable threat intelligence (TI) reports. We present TIBlender, a multi-agent system that monitors four platforms (X, Reddit, Telegram, and Discord) and produces structured TI reports via role-specialized LLM agents. These agents conduct multi-perspective investigations, tracing chains of evidence to uncover related Indicators of Compromise (IoCs) via collaborative, evidence-backed analysis. In a real-world deployment, TIBlender detected emerging threats across all four threat categories ahead of public feeds, including in-the-wild exploitation ahead of public vulnerability registries; the majority of its IoCs were absent from each evaluated feed. Quantitative evaluation confirms that each platform contributes unique threat information unavailable from the others, and that excluding any single platform results in substantial loss of reports in specific threat categories. Under identical single-platform input conditions, TIBlender's direct extraction produced substantially different IoC sets from the corresponding platform-specific baselines, while multi-perspective investigation further enriched the resulting intelligence. These results establish cross-platform social media monitoring as an effective and scalable early-warning layer for operational TI pipelines.
\end{abstract}

\begin{IEEEkeywords}
Threat Intelligence, Multi-Agent System, Large Language Model, Social Media
\end{IEEEkeywords}

\section{Introduction}
\label{sec:introduction}

As cyber attacks become increasingly sophisticated, the rapid collection and operationalization of threat intelligence (TI) have become critical to organizational defense.
Conventional TI collection relies heavily on commercial and public threat feeds and databases, introducing an inevitable time lag between the emergence of a new threat and its collection, validation, and publication~\cite{DBLP:conf/ccs/BouwmanEHGE25,galloway2026actively}.
In contrast, social media platforms have been shown to carry Indicators of Compromise (IoCs) and attack campaign details ahead of their publication in official threat feeds and databases~\cite{DBLP:conf/www/ShinSKLKH21,DBLP:conf/raid/PaladiniFPZC24}, drawing attention to these platforms as early-warning channels for TI collection.

These platforms host communities with diverse roles and motivations.
Security researchers share IoCs, Proofs of Concept (PoCs), and vulnerability disclosures on Reddit and X as community warnings; adversaries advertise and sell malware and phishing kits on Telegram; and victims report attacks on Discord and Reddit.
In other words, signals from the same attack campaign (a coordinated body of threat activity sharing common targets, attack vectors, and threat types) are distributed in fragmented, multilingual form across multiple platforms.
Building on the observation that social media platforms carry threat signals ahead of official publication, prior work has studied automated IoC extraction~\cite{DBLP:conf/ndss/CuiKJYKLC0L25,DBLP:conf/bigdataconf/NiakanlahijiSHC19} and the automation of threat analysis using Large Language Models (LLMs)~\cite{DBLP:journals/corr/abs-2508-10677,DBLP:conf/uss/BuchelPLCZBE0GC25}.
However, these approaches either target a specific platform or address individual tasks such as IoC extraction or threat detection.
None has achieved end-to-end automation that integrates multilingual information across multiple platforms and generates actionable TI reports.
Realizing this capability requires overcoming the following three fundamental challenges.

\begin{figure}[t!]
 \centering
 \includegraphics[width=0.48\textwidth]{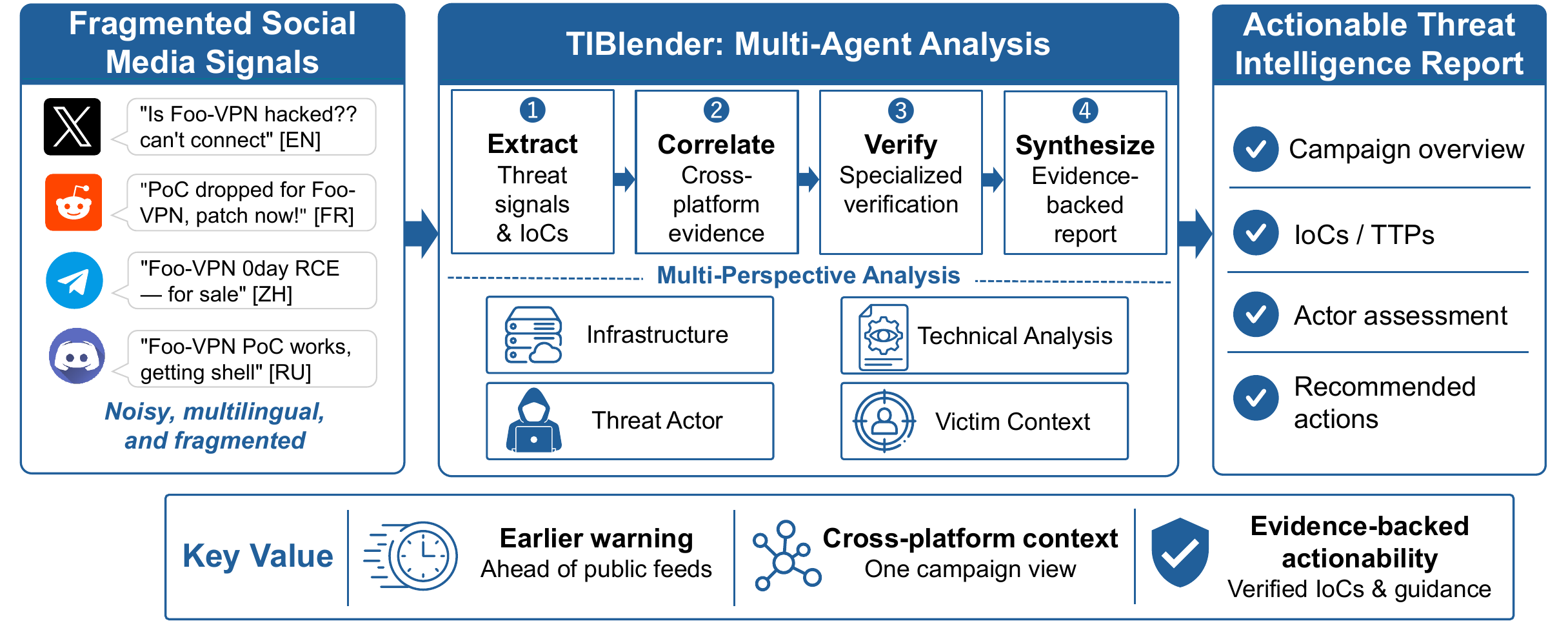}
 \caption{TIBlender concept overview.
Illustrative posts are shown in English; language tags
(EN/FR/ZH/RU) indicate their source languages.
TIBlender integrates fragmented social media signals
into actionable TI reports through multi-agent analysis.}
\label{fig:concept}
\end{figure}

\noindent\textbf{First, integrating fragmented, multilingual information across heterogeneous
platforms.}
Threat-related information is shared in fragmented form across multiple platforms, each
capturing a different aspect of a threat: infrastructure, attack techniques, victim impact,
or threat actor
activity~\cite{DBLP:conf/datamod/JakstaiteC23,DBLP:conf/raid/PaladiniFPZC24}.
Since each platform exposes only one aspect of a threat, monitoring a single platform yields
intelligence that is structurally incomplete.
Moreover, platforms differ in language, posting culture, and information granularity;
correlating heterogeneous, multilingual, multi-format signals to derive comprehensive threat
intelligence remains unaddressed by existing single-platform approaches.

\noindent\textbf{Second, validating the reliability of automatically extracted threat
information.}
Social media posts contain a mix of noise, misinformation, and exaggeration, and automated
extraction relying on a single perspective cannot fully eliminate false
positives~\cite{DBLP:conf/IEEEares/MezziMT25}.
Integrating TI directly into security operations requires cross-validating evidence from
multiple independent perspectives while keeping each decision traceable to its supporting
evidence.
This capability is absent from existing approaches~\cite{DBLP:journals/compsec/TounsiR18}.

\noindent\textbf{Third, filtering massive noise while adapting to a shifting threat
environment.}
Threat signals, including IoCs and vulnerability disclosures shared by security researchers and attack reports posted by victims, are buried under chatter, advertisements, and spam~\cite{DBLP:conf/raid/AlamBPR23}.
Efficiently separating these signals from surrounding noise is therefore essential.
Furthermore, since the targets, techniques, and platforms used by adversaries constantly
evolve, static monitoring rules and queries cannot keep pace with emerging threats.

\emph{No existing system simultaneously addresses all three challenges: integrating fragmented multi-platform data, validating reliability across sources, and filtering noise while adapting to an evolving threat landscape.}

To address these challenges, we propose TIBlender, an autonomous multi-agent system (Figure~\ref{fig:concept}).
TIBlender is designed to be platform-agnostic; in this work, we deploy and evaluate it across four representative platforms (X, Reddit, Telegram, and Discord) selected for their complementary roles as TI sources (Section~\ref{subsec:scope}).
It monitors these platforms to identify and consolidate related threat signals into campaign-level clusters, then deploys investigation agents covering four perspectives (Infrastructure, Technical, Social, and Actor) to conduct parallel investigations using external tools such as RDAP/WHOIS, Passive DNS, and CVE databases.
Multiple independent evaluation agents cross-validate the investigation results.
Only clusters with sufficient evidence are released as TI reports for direct use in Security
Operations Centers (SOC) or Security Information and Event Management
(SIEM) systems~\cite{DBLP:journals/sensors/GranadilloZD21}.
In addition, Reinforcement Learning (RL)-Guided investigation and trend-based query adaptation continuously improve IoC yield per investigation cycle while adapting to shifts in the threat environment.

Over a 31-day deployment spanning four platforms, TIBlender autonomously generated
8,288 actionable threat reports.
Excluding any single platform eliminates up to 50\% of reports in specific threat categories,
and over 80\% of TIBlender's directly extracted IoCs were absent from the output of single-platform baselines under identical input conditions.
Across six evaluated public feeds, 83.0\%--99.6\% of TIBlender's IoCs were absent from each; TIBlender captured in-the-wild exploitation up to 13.4 days before CISA KEV listing~\cite{cisa_kev}.
The ablation study confirms that each design component individually contributes to
detection performance.
 The full source code, generated reports, and supporting evaluation artifacts are publicly available.\footnote{\url{https://tiblender.github.io/}}

\medskip
\noindent\textbf{Our contributions are as follows.}

\begin{itemize}
\item We propose and implement TIBlender, a multi-agent system that integrates
 threat signals fragmented across multiple platforms and autonomously generates actionable
 TI reports with verified IoCs and clear, evidence-backed justifications.

\item Through quantitative evaluation on real-world operational data, we show that each
 platform provides unique threat information unavailable from the others, and that
 cross-platform integration is essential for comprehensive TI coverage.

\item Through large-scale operational deployment across four platforms, we demonstrate
 that TIBlender can capture threats ahead of public threat feeds and public vulnerability
 registries.
\end{itemize}

\section{Background}
\label{sec:background}

This section reviews prior work related to TIBlender's design, positions our approach
relative to existing work, and defines the scope of this study.

\subsection{Related Work}
\label{subsec:related_work}

\noindent\textbf{Online communities as sources of early threat intelligence.}
Social media platforms and underground forums have been shown to carry threat signals
ahead of public feeds~\cite{DBLP:conf/raid/0017YL0Z20,DBLP:conf/acl/JinJCCLS23,DBLP:journals/fgcs/CaballeroGMSSV23,DBLP:conf/IEEEares/Nakano0KFYHYM23}.
CVE mentions on X (formerly known as Twitter) were first quantified as early exploitation
indicators~\cite{DBLP:conf/uss/SabottkeSD15}; a large-scale analysis of threat data on
X revealed time-lag characteristics relative to public
feeds~\cite{DBLP:conf/esorics/AlvesAGFB20}; and TwiTi confirmed that IoCs on X precede
public feeds by several days~\cite{DBLP:conf/www/ShinSKLKH21}.
A comparison of CVE mention patterns on X and Reddit found that the two platforms surface
threats at different speeds and depths~\cite{DBLP:conf/webi/Horawalavithana19}.
Threats discussed on underground forums appear ahead of official security reports in over 60\% of cases~\cite{DBLP:conf/raid/PaladiniFPZC24}, and deep/dark web analysis has characterized
attacker-side supply chains in detail~\cite{DBLP:conf/www/KimZKH26}.
These findings collectively show that early threat dissemination spans multiple community types.
\emph{In contrast to retrospective ecosystem analysis, our work focuses on real-time
monitoring of publicly accessible social media platforms to detect emerging threats
before they appear in public feeds.}

\noindent\textbf{Platform-specific threat detection and its limitations.}
Threat detection and IoC extraction on individual platforms have been widely
studied~\cite{DBLP:conf/esorics/LiZCL22,DBLP:conf/ndss/LiL24,DBLP:conf/ccs/QinXL23,DBLP:conf/ccs/Bouma-SimsLC25}.
Tweezers detects cyber events on X via BERT-based classification and knowledge graph
augmentation~\cite{DBLP:conf/ndss/CuiKJYKLC0L25}; DarkGram detects malicious posts on
Telegram at scale~\cite{DBLP:conf/uss/RoyVKN25}; and adversarial use of Discord has also
been documented~\cite{intel471_discord_2023}.
IoC extraction has also advanced through contextual methods
(IoCMiner~\cite{DBLP:conf/bigdataconf/NiakanlahijiSHC19}), OSINT-based
collection~\cite{liao2016acing}, multi-stage
pipelines~\cite{DBLP:conf/datamod/JakstaiteC23}, domain-tagged CTI generation from
social data (TIMiner~\cite{DBLP:journals/compsec/ZhaoYLSHL20}), early IoC detection on
underground forums (IoC Stalker~\cite{DBLP:conf/acsac/MischingerPS24}), and automated
report collection with benchmarked IoC extraction
(GoodFATR~\cite{DBLP:journals/fgcs/CaballeroGMSSV23}).
\emph{All of these are limited to a specific platform or individual tasks; none addresses
cross-platform integration or autonomous report generation.}

\noindent\textbf{Automated TI Analysis and LLM Applications.}
Automated methods for TI structuring have advanced from
NLP-based extraction to LLM-driven approaches~\cite{DBLP:conf/icde/GaoSCXLLZ21,DBLP:journals/compsec/HuZHSW24,DBLP:journals/compsec/ZhangDMWXYLC25}:
TTPDrill extracts Tactics, Techniques, and Procedures (TTPs) and maps them to MITRE
ATT\&CK~\cite{strom2018attck,DBLP:conf/acsac/HusariAACN17}; ChainSmith assigns semantic
roles to IoCs within attack campaigns~\cite{DBLP:conf/eurosp/ZhuD18}; and CTINEXUS
constructs cybersecurity knowledge graphs via in-context
learning~\cite{DBLP:conf/eurosp/ChengBTSG25}.
However, LLMs have been shown to be inconsistent on full-length Advanced Persistent Threat
(APT) reports~\cite{DBLP:conf/IEEEares/MezziMT25}, and TTP extraction methods involve inherent precision-recall trade-offs~\cite{DBLP:conf/uss/BuchelPLCZBE0GC25}.
\emph{All of these methods assume structured TI reports as input; none is designed for fragmented, multilingual social media data.}

\subsection{Research Scope}
\label{subsec:scope}

Prior work establishes social media as a rich and timely TI source,
yet no existing system autonomously generates actionable TI from
fragmented, multilingual, cross-platform data---a gap that, alongside
those identified in Section~\ref{sec:introduction}, defines the scope of this study.

\noindent\textbf{Target platforms and their characteristics.}
We continuously monitor four platforms encompassing both fully public and
semi-public community structures; their primary characteristics are summarized
in Table~\ref{tab:platforms}.

 Prior work has shown that IoC sharing on X and Reddit can precede public
feeds~\cite{DBLP:conf/www/ShinSKLKH21,DBLP:conf/webi/Horawalavithana19}.
X shows the strongest tendency for first IoC reports to precede public
feeds~\cite{DBLP:conf/www/ShinSKLKH21}; Reddit's security-focused subreddits
 (e.g., r/netsec, r/malware) host in-depth technical discussions of PoCs,
vulnerability analyses, and attack techniques~\cite{DBLP:conf/webi/Horawalavithana19};
Telegram's semi-public channels are a documented venue for threat actors sharing
 C2 infrastructure details, malware distribution, and stolen-data
 sales~\cite{DBLP:conf/uss/RoyVKN25}; and Discord has seen growing threat-actor use
 for coordination and technical information sharing~\cite{intel471_discord_2023}.

 Collection is limited to posts from public and semi-public communities
 (private messages and direct messages are excluded), and all languages are in scope.

\begin{table}[t!]
\scriptsize
\centering
\caption{Information characteristics of the monitored platforms}
\label{tab:platforms}
{\setlength{\tabcolsep}{4pt}
\begin{tabular}{@{}llll@{}}
\toprule
\textbf{Platform} & \textbf{Openness} & \textbf{Primary Characteristics} & \textbf{Role as TI Source} \\
\midrule
X & Public & Virality & Breaking IoC Reports \\
Reddit & Public & In-depth discussion & Technical Analysis \\
Telegram & Semi-public & Anonymity & Threat Actor Activity \\
Discord & Semi-public & Private community & Actor Coordination \\
\bottomrule
\end{tabular}}
\end{table}

\noindent\textbf{Target threat categories and IoC types.}
We target four threat categories: Phishing/Fraud, Malware/C2, Vulnerability, and
APT/Ransomware, extracting IoCs such as URLs, file hashes, domains, IP addresses, and CVE
identifiers from each category.
These four categories cover the major threats that are predominantly discussed on social media.
Evaluation metrics and measurement methods for each category are described in
Section~\ref{sec:setup}.

\section{TIBlender System Design}
\label{sec:system}

This section describes the design of TIBlender's five-step in-cycle pipeline and the
cross-cycle feedback mechanisms.

\subsection{System Overview}
\label{subsec:overview}

\begin{figure*}[t]
\centering
\includegraphics[width=\textwidth]{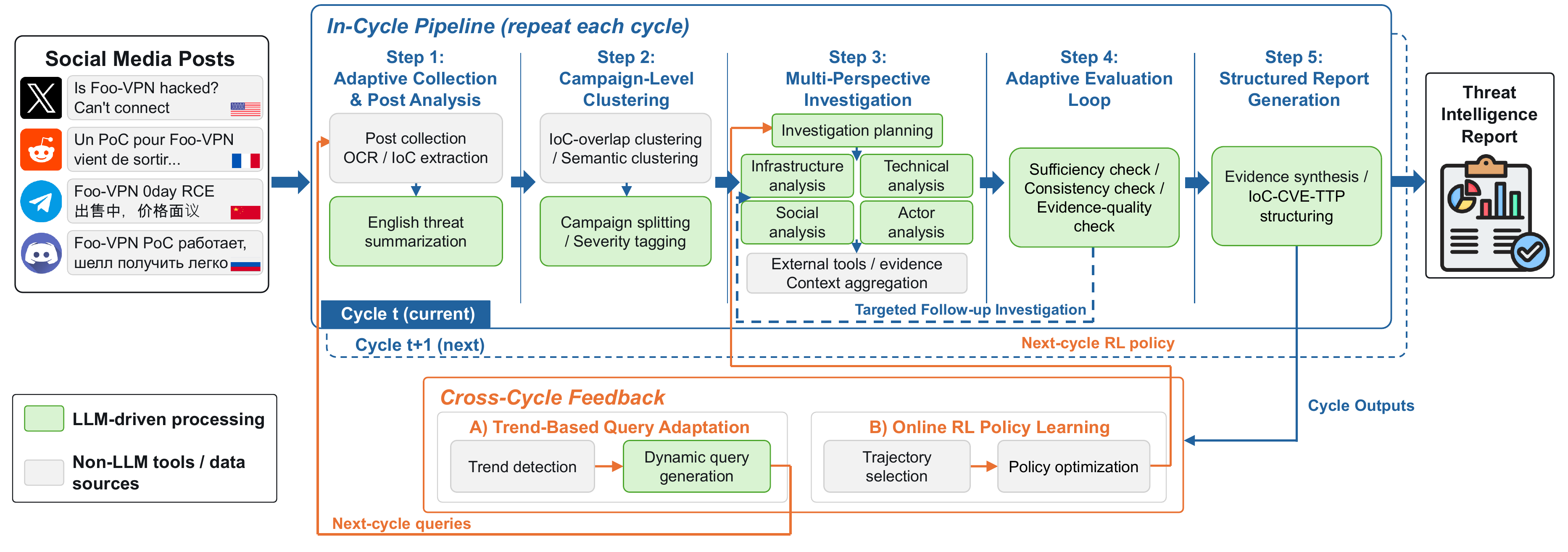}
\caption{System architecture of TIBlender. Solid arrows represent the five-step in-cycle
pipeline that transforms social media posts into structured threat intelligence reports.
The dashed arrow from Step~4 to Step~3 represents the evaluation loop, which triggers
follow-up investigation when evidence quality is insufficient. The two cross-cycle feedback
mechanisms at the bottom take \textit{Cycle Outputs} as input to update dynamic queries for
Step~1 and the RL policy for Step~3 of the next cycle.}
\label{fig:architecture}
\end{figure*}

TIBlender is a multi-agent system that integrates fragmented threat information from social
media and autonomously generates threat intelligence reports with immediately deployable
IoCs and explicit evidence-backed justifications.
Figure~\ref{fig:architecture} shows the overall architecture.

TIBlender operates in fixed-interval cycles, each processing posts collected
during the immediately preceding time window.
It then sequentially executes the five-step in-cycle pipeline: Step~1 (Adaptive Collection
and Post Analysis), Step~2 (Campaign-Level Clustering), Step~3 (Multi-Perspective
Investigation), Step~4 (Adaptive Evaluation Loop), and Step~5 (Structured Report
Generation).
Outputs are produced as TI reports in JSON, PDF, Markdown, and STIX~2.1~\cite{oasis2021stix}
formats; cycle interval and time window values are given in Section~\ref{sec:setup}.
Beyond this main pipeline, an evaluation loop within Step~4 triggers follow-up
investigation in Step~3 when evidence quality is deemed insufficient.
Additionally, at the end of every cycle, the generated reports and investigation trajectories
(\textit{Cycle Outputs}) are fed into the trend-based query adaptation and the online RL
policy learning, which update Step~1 and Step~3 of the next cycle.
Each step is driven by specialized LLM agents, each defined by a role-specific prompt and
input/output schema, ranging from those that invoke external tools to those that perform
structured data transformations.

\subsection{Step 1: Adaptive Collection and Post Analysis}
\label{subsec:step1}

Step~1 adaptively collects threat-related posts from four platforms and converts raw posts
in diverse languages and formats into (threat summary, IoC list) pairs that downstream
clustering and investigation steps can process uniformly.

\noindent\textbf{Post collection.}
Collection combines two mechanisms: continuous monitoring using predefined multilingual keywords, and dynamic keyword augmentation generated by the trend-based query adaptation (Section~\ref{subsec:cross-cycle-feedback}).
Dynamic keyword augmentation extends coverage to emerging threats (keyword counts and supported languages are detailed in Appendix~\ref{appendix:params}).

For X, predefined keywords and designated accounts are monitored continuously; dynamic
queries extend coverage via keyword search.
For Reddit, posts from predefined security-focused subreddits are monitored continuously;
dynamic queries supplement coverage through keyword searches across all subreddits.
For Telegram, seed channels are identified via in-platform keyword search; related public
 channels are then progressively discovered and added by following channel links and
recommendations starting from those seeds.
For Discord, invitation URL candidates are automatically collected via the Discord
server discovery API, web search engines, and server directory sites; an analyst
screens and manually joins selected servers, after which post retrieval and analysis
 proceed automatically. The access interfaces and, where applicable, joining criteria
are detailed in Section~\ref{subsec:ethics}.

\noindent\textbf{Preprocessing.}
After collection, platform-specific rule-based filters are applied to remove obvious noise
before downstream processing.
Short posts containing no security-relevant keywords such as malware names,
CVE identifiers, or IoCs are discarded; all other posts proceed to downstream processing.
TIBlender then applies Optical Character Recognition (OCR) to images attached to each post and merges the extracted text into that post's body for subsequent processing.

\noindent\textbf{Threat summarization and IoC extraction.}
Using the merged text as input, an LLM agent generates an English threat summary while
regex-based IoC extraction (URLs, file hashes, domains, IP addresses, and CVE identifiers)
runs in parallel.
English summarization normalizes multilingual posts into a common language so that
downstream clustering is unaffected by linguistic variation.
Each post yields a (threat summary, IoC list) pair passed to subsequent steps; for posts
containing no IoCs, the list is empty.
Posts without IoCs are retained because they may contain important contextual information
about attack background and tactics that contributes to downstream investigation.

\subsection{Step 2: Campaign-Level Clustering}
\label{subsec:step2}

Step~2 consolidates posts scattered across four platforms into clusters
corresponding to the same threat campaign, shifting the unit of processing
from the post level to the campaign level for all downstream steps.
Using the (threat summary, IoC list) pairs from Step~1 as input, TIBlender forms clusters in
two stages: clustering via two complementary signals (semantic similarity and IoC overlap),
followed by LLM-based campaign splitting.
Using IoC overlap alone risks missing posts that contain no IoCs; using semantic
similarity alone risks incorrectly merging posts from different campaigns that happen
to use similar wording.
TIBlender therefore combines both signals, directly addressing each failure mode.

\noindent\textbf{Clustering.}
Posts are split into two tracks according to whether they contain IoCs, with each signal applied to the cases where it is most reliable: the IoC track
uses IoC overlap for precise linking, while the non-IoC track uses semantic
similarity to attach posts to existing IoC-track clusters.
In the IoC track, post pairs whose IoC Jaccard similarity meets or exceeds the threshold $\tau_J$ are
iteratively merged using the union-find algorithm, forming connected components as IoC-track clusters.
In the non-IoC track, embedding vectors of threat summaries generated in Step~1 are used to
absorb posts whose cosine similarity to the centroid of the nearest IoC-track cluster meets
the absorption threshold $\tau_{\text{abs}}$; remaining unabsorbed posts are grouped into independent clusters
using HDBSCAN~\cite{campello2015hdbscan} (see Appendix~\ref{appendix:params} for models and thresholds).

\noindent\textbf{LLM-based campaign splitting and severity tagging.}
While clustering by semantic similarity and IoC overlap effectively consolidates posts from
the same campaign, posts from distinct campaigns with similar infrastructure or techniques
may end up in the same cluster.
Since similarity signals alone cannot distinguish such cases, an LLM agent autonomously
splits each cluster into campaign-level units based on target organization, attack vector,
and threat type.
It then assigns each unit a topic name and severity level (LOW/MEDIUM/HIGH/CRITICAL).
Only clusters meeting the configurable severity threshold $\theta_{\text{sev}}$ proceed to
Step~3 (values in Section~\ref{subsec:impl-details}).
To prevent misclassification due to campaign mixing, we adopt a conservative strategy that
tolerates over-splitting.

\subsection{Step 3: Multi-Perspective Investigation}
\label{subsec:step3}

For each threat campaign, Step~3 assembles the supporting evidence needed
for downstream report generation by collecting and organizing findings
across four specialized perspectives (Infrastructure, Technical, Social,
and Actor).
For each cluster with severity $\geq \theta_{\text{sev}}$ from Step~2, TIBlender collects
and organizes evidence through three phases: pre-investigation preparation, parallel
investigation by four perspective agents, and contextual aggregation before passing to Step~4.
The parallel investigation phase operates in one of two modes (Multi-Perspective or RL-Guided) selected based on similarity to past cases, as described below.
\begin{table}[!tbp]
\scriptsize
\centering
\caption{The four investigation perspectives and their investigation focuses in Step~3.}
\label{tab:perspectives}
\begin{tabular}{@{}p{2cm}p{6.2cm}@{}}
\toprule
\textbf{Perspective} & \textbf{Investigation Focuses (18 Total)} \\
\midrule
Infrastructure \newline (Where It Operates) & domain registration and RDAP/WHOIS lookup, DNS and Passive DNS analysis, TLS certificate analysis, related domain and infrastructure pivoting, hosting and OSINT collection \\
\midrule
Technical \newline (What It Does) & CVE exploitation analysis, PoC and exploit code investigation, related webpage and technical document retrieval, technical intelligence and security report collection \\
\midrule
Social \newline (How It Spreads) & phishing campaign tracking, social engineering tactics analysis, brand impersonation and spoofing investigation, SNS account abuse and spread tracking \\
\midrule
Actor \newline (Who Is Behind It) & threat actor attribution and profiling, TTP mapping, campaign tracking, APT group and threat actor investigation, cybercrime group and market analysis \\
\bottomrule
\end{tabular}
\end{table}

\noindent\textbf{Investigation mode selection.}
The investigation mode for each cluster is determined by its similarity score to TI reports
accumulated in past cycles.
Candidate reports are retrieved via vector database search, and similarity is computed by a
trained scoring model using features such as IoC overlap, CVE overlap, and semantic
similarity (details in Appendix~\ref{appendix:params}).
Clusters below the similarity threshold $\theta_{\text{rl}}$ are processed in
\textbf{(1)~Multi-Perspective mode} (unknown or novel threats); those at or above
$\theta_{\text{rl}}$ use \textbf{(2)~RL-Guided mode} (threats whose IoCs, CVEs, and
topic resemble past cases).

\noindent\textbf{(1)~Multi-Perspective mode.}
Prior to the parallel investigation, a dedicated planning agent takes
a cluster-level overview and generates a prioritized query plan spanning all four
perspectives, which is passed to each perspective agent as input.
Four LLM agents (one per perspective: Infrastructure, Technical, Social, and Actor; 
Table~\ref{tab:perspectives}) run in parallel, each independently executing 
external tools to collect evidence according to the strategy plan.
Each agent performs chained pivots starting from IoCs or domains found in the
initial investigation, using tools such as Passive DNS, RDAP/WHOIS, and search engines
to track related infrastructure and threat actor information.

\noindent\textbf{(2)~RL-Guided mode.}
RL-Guided mode uses the same four perspective LLM agents as Multi-Perspective mode, but an
RL policy predetermines the priority allocation of perspectives and investigation focuses
and passes them as strategic hints to each agent
(the state space, action space, and reward function are formalized in
Appendix~\ref{appendix:rl}).
For example, a perspective that historically proved effective for a cluster's threat
type (e.g., Infrastructure / domain analysis for phishing) is assigned higher priority.
During the cold-start phase (early deployment when accumulated reports are scarce),
similarity scores rarely exceed $\theta_{\text{rl}}$, so the system runs in Multi-Perspective mode;
as more cases accumulate, the proportion of clusters processed in RL-Guided mode increases.

\noindent\textbf{IoC extraction type and confidence.}
IoCs collected in both modes are classified by extraction type:
those extracted directly from cluster posts are classified as \textsc{Direct}; those
discovered by pivoting from known IoCs to related infrastructure via external tools are
classified as \textsc{Pivot}.
A three-level confidence score is assigned using rules based on the number of independent
sources, shared hosting detection, and cross-tool consistency (details in
Appendix~\ref{appendix:params}).
Confidence scores determine the processing priority of each IoC in Step~4's quality evaluation and Step~5's report generation.

\noindent\textbf{Context aggregation.}
After all perspectives complete their investigations, a context aggregation step consolidates
the evidence across perspectives and filters out benign indicators, including CDN domains and legitimate service domains, that are unlikely to represent malicious infrastructure.
Each piece of evidence is tagged with the perspective that collected it (Infrastructure, Technical, Social, or Actor), confidence level, and the tool used.
It is then passed to Step~4 as an investigation context containing per-perspective IoCs,
CVEs, and TTPs along with each perspective's investigation trajectory.
The trajectory records all tool calls and evidence collected during that perspective's investigation.
This per-perspective tagging allows Step~4's evaluation agents to identify which
perspectives have insufficient evidence and request targeted follow-up investigation.

\subsection{Step 4: Adaptive Evaluation Loop}
\label{subsec:step4}

Using judge agents independent of the investigation agents, Step~4 verifies
the quality of evidence collected in Step~3 and passes only clusters
meeting quality criteria to report generation.

\noindent\textbf{Three-axis quality evaluation.}
Two judges assess the investigation context along three axes.
Sufficiency assesses whether IoCs, CVEs, and TTPs required for threat analysis
have been adequately collected across all investigation perspectives.
Consistency detects contradictions among evidence collected from multiple
perspectives (e.g., conflicting attributions for the same IP address).
Based on the IoC extraction type and confidence scores assigned in Step~3,
evidence quality verifies that HIGH-confidence evidence is sufficiently represented
and that the evidence base is not dominated by LOW-confidence IoCs.

\noindent\textbf{Multi-judge evaluation loop.}
To reduce dependence on a single model, two judges from different LLM providers
(with different training data and alignment policies) evaluate the same
investigation context in parallel without observing each other's outputs,
thereby reducing the risk of model-specific systematic
bias~\cite{wang2023selfconsistency}.
Each judge outputs one of three verdicts: \texttt{PROCEED},
\texttt{NEED\_MORE\_INFO}, or \texttt{SKIP}.
When a judge returns \texttt{NEED\_MORE\_INFO}, it generates follow-up queries
targeting insufficient perspectives, and Step~3 is re-invoked in
Multi-Perspective mode (RL mode is not used for follow-up).
The resulting evidence is merged and re-evaluated.
The final verdict uses conservative consensus: \texttt{PROCEED} is returned
only when all judges return \texttt{PROCEED}; if any judge returns \texttt{SKIP},
or the iteration limit $N_{\max}$ is reached without consensus, the overall
verdict is \texttt{SKIP}.

Since false positives directly increase analyst workload in operational TI
settings, we explicitly accept a modest increase in false negative rates as a
design trade-off in favor of false positive suppression; the resulting FN rate
is quantified in Section~\ref{subsec:human-eval-design}.

\subsection{Step 5: Structured Report Generation}
\label{subsec:step5}

Step~5 converts the investigation context that has passed Step~4's
evaluation into a structured TI report with immediately deployable IoCs
and explicit justifications.
The resulting report is exported in formats suitable for both manual analyst review and
automated integration with SIEM and Security Orchestration, Automation and Response
(SOAR) systems.

The report generation agent fills in each field sequentially using prompts specific to each section.
Structured fields for IoCs and TTPs are populated via LLM-based JSON extraction.
Free-text fields such as Executive Summary and Technical Analysis are handled as generation
tasks referencing the full context.
All free-text fields are generated in English, with explicit justifications (including
supporting evidence, confidence rationale, and any detected contradictions) embedded in the
report.
The output is a structured JSON with 11 sections: cluster name, severity
(LOW/MEDIUM/HIGH/CRITICAL), executive summary, key findings, recommendations,
IoCs (with confidence and extraction type), CVEs, TTPs, threat actor, attack timeline,
and source evidence.
Conversion to PDF, Markdown, and STIX~2.1~\cite{oasis2021stix} formats is also supported.
Figure~\ref{fig:report-example} shows an excerpt from a TI report generated during
TIBlender's operational deployment, adapted for readability; the full report is provided
in Appendix~\ref{appendix:report-full}.

\begin{figure}[!tbp]
\begin{tcolorbox}[
 title={\textbf{Adobe Reader Zero-Day PDF Exploitation (Excerpt)}},
 colback=gray!4,
 colframe=black!60,
 fonttitle={\bfseries\scriptsize},
 fontupper=\scriptsize,
 left=4pt,
 right=4pt,
 top=3pt,
 bottom=3pt
]
\textbf{threat\_severity:}~\texttt{CRITICAL} \hfill

\textbf{Summary:}~\textit{An unpatched zero-day in Adobe Reader (no CVE assigned) has been exploited in the wild since December 2025 via maliciously crafted PDFs. Obfuscated JavaScript abuses a privileged Acrobat API to fingerprint the system, exfiltrate data, and fetch staged payloads from attacker-controlled servers. Russian-language lures referencing current events in Russia's oil \& gas sector were reported, suggesting possible targeting.}\\[3pt]

{\setlength\tabcolsep{4pt}%
\begin{tabularx}{\linewidth}{@{}lllX@{}}
\toprule
\textbf{Action} & \textbf{Type} & \textbf{Indicator} & \textbf{Rationale}\\
\midrule
\textsc{block} & domain & \texttt{ado-read-parser[.]com} & C2 domain\\
\textsc{block} & ipv4 & \texttt{169.40.2[.]68} & C2 server\\
\textsc{block} & ipv4 & \texttt{188.214.34[.]20} & Payload staging server\\
\textsc{block} & sha256 & \texttt{65dca34b{\dots}de7} & Malicious PDF sample\\
\bottomrule
\end{tabularx}}\\[3pt]

\textbf{TTPs:}~T1204.002 (Malicious File), T1059.007 (JavaScript), T1041 (Exfiltration over C2)\\[3pt]

{\setlength\tabcolsep{4pt}%
\begin{tabularx}{\linewidth}{@{}lX@{}}
\toprule
\textbf{Platform} & \textbf{Unique contribution}\\
\midrule
X & C2 domain \texttt{ado-read-parser[.]com} and global exploitation alerts, including international-language coverage confirming active spread\\
Discord & Exploitation mechanism: zero-day observed in the wild via JS Engine\\
Reddit & Attack timeline (active since December 2025); oil \& gas sector targeting with Russian-language lures; malicious PDF artifact traced to researcher analysis, yielding SHA-256 hash\\
\bottomrule
\end{tabularx}}
\end{tcolorbox}

{\scriptsize
\caption{TIBlender TI report excerpt for an Adobe
Reader zero-day (later CVE-2026-34621),
integrating X (C2 and alerts), Discord
(exploitation vector), and Reddit
(timeline and malware artifact).
Generated 1 day 9 hours before CVE publication
and 3 days 17 hours before public-feed listing.
Full report: Appendix~\ref{appendix:report-full}.}
\label{fig:report-example}
}
\end{figure}

\subsection{Cross-Cycle Feedback}
\label{subsec:cross-cycle-feedback}

The \textit{cross-cycle feedback} takes each cycle's \textit{Cycle Outputs} 
(Step~5 reports and Step~3 investigation trajectories, i.e., records of 
agent actions and collected evidence) as input to update Step~1's collection 
queries and Step~3's RL policy for the next cycle.

\noindent\textbf{Trend-based query adaptation.}
Static monitoring queries introduce delays in responding to emerging threats and may
favor certain languages and geographic regions, missing signals that arise in
underrepresented linguistic communities.
This mechanism compares the current cycle's cluster set against a
rolling historical snapshot of clusters accumulated over the past $W$ hours
to identify emerging trends (adjacency criterion: $\text{sim}(c,h) \geq \tau_{\text{sim}}$).
Four trend detection axes run in parallel: \textsc{Volume}, \textsc{Propagation}, \textsc{Escalation}, and \textsc{Early\_Warning}.
\textsc{Volume} flags clusters whose per-cycle occurrence count exceeds $\mu + k\sigma$, where $\mu$ and $\sigma$ are the mean and standard deviation of historical cluster counts per cycle.
\textsc{Propagation} tracks cross-platform spread, \textsc{Escalation} detects rising severity, and \textsc{Early\_Warning} flags clusters whose LLM-assessed strategic value score meets or exceeds $\theta_{\text{ew}}$.
Clusters satisfying at least one criterion are added to the trend set
(threshold details in Appendix~\ref{appendix:params}).
The context of each detected trend type is forwarded to an LLM to generate dynamic queries
adapted to the threat's language, geographic scope, and context, which are then passed to
Step~1's collection mechanism for the next cycle.

\noindent\textbf{Online RL policy learning.}
At the end of each cycle, investigation trajectories from
the past $W$ hours are collected as $(s_t,a_t,r_t)$ tuples
(Equation~(\ref{eq:reward})).
Only trajectories meeting the quality criterion
(severity $\geq$ MEDIUM) are retained.
Once a minimum number has accumulated, the policy is
updated each cycle through behavior cloning (BC) followed
by Deep Q-Network (DQN) fine-tuning.
BC learns from successful investigation demonstrations,
while DQN uses reward feedback for discrete perspective
allocation, balancing adaptation and learning stability.
The policy initially learns from Multi-Perspective
trajectories; as RL-Guided trajectories accumulate,
their contribution increases, improving subsequent
investigation strategies (Appendix~\ref{appendix:rl}).

\section{Evaluation}
\label{sec:setup}

This section evaluates TIBlender across five axes: (1) cross-platform monitoring effectiveness, (2) same-input comparison with IoC extraction baselines, (3) comparison with public threat feeds, (4) human validation and filtering analysis, and (5) ablation study; case studies are in Section~\ref{subsec:case-studies}.

\subsection{Deployment Setup and Implementation Parameters}
\label{subsec:deployment}

We deployed TIBlender across four platforms (X, Reddit, Telegram, and Discord) from
March~16 to April~16, 2026, for a continuous 31-day operation.
Collection cycles ran hourly, totaling 744 cycles over 31 days.
Prior to this evaluation period, the RL policy was pre-trained on
Multi-Perspective investigation trajectories during a one-week warm-up period
(March~9--15, 2026); details are provided in Appendix~\ref{appendix:rl}.

\noindent\textbf{Computing environment.}
The system ran on Azure Standard D32as v4 (32 vCPUs, 128\,GB RAM) with Ubuntu 22.04, with
collection, analysis, and report generation modules orchestrated via Docker Compose.

\noindent\textbf{LLM configuration.}
All models were accessed via Microsoft Foundry.
Grok 4 Fast Non-Reasoning (xAI) was used for normalization, clustering, multi-perspective
investigation, report generation, trend detection, and query generation.
This model was selected for its strong performance on short-form, multilingual text.
GPT-OSS-120B (OpenAI) and Llama 4 Maverick (Meta) were used in parallel for the adaptive
evaluation phase, for the same bias-reduction rationale as Step~4 (Section~\ref{subsec:step4}).
All models were run at temperature 0.0 for reproducibility.

\noindent\textbf{Implementation parameters.}
\label{subsec:impl-details}
Parameters were determined through a combination of design constraints and preliminary experiments (detailed rationale in Appendix~\ref{appendix:params}).
Step~1 collection keywords comprised approximately 100 terms across four threat categories
in nine languages.
Step~2 clustering uses HDBSCAN, which requires no prior specification of cluster count
($min\_cluster\_size=3$, IoC overlap threshold $\tau_J=0.25$, absorption threshold
$\tau_{\text{abs}}=0.75$); only clusters with severity $\geq$ MEDIUM
proceed to Step~3.
Step~3 switches between Multi-Perspective and RL-Guided modes via a
similarity threshold ($\theta_{\text{rl}}=0.60$).
Step~4's adaptive evaluation loop converges within at most $N_{\max}=3$ iterations.
Trend detection operates with a snapshot window of $W=168$ hours, similarity threshold
$\tau_{\text{sim}}=0.80$, volume anomaly coefficient $k=1.5$, and early warning threshold
$\theta_{\text{ew}}=0.75$.

\subsection{31-Day Operational Overview}
\label{subsec:operational-utility}

\noindent\textbf{Collection scale and clustering results.}
A total of 873,973 raw posts were collected from four platforms over 31 days
(Table~\ref{tab:dataset-overview}).
Telegram and Discord, collected exhaustively, contained high volumes of everyday
conversation and spam, whereas X and Reddit, collected via keyword filtering, had higher
threat-relevant density.
These 873,973 raw posts were consolidated into 184,240 unique clusters, and after
successive filtering stages, 8,288 actionable threat reports were generated
(Table~\ref{tab:dataset-overview}).
The primary filters were LOW-severity exclusion at Step~2 (172,683 clusters,
93.7\%) and \texttt{SKIP} verdicts at Step~4 (3,269 clusters, 1.8\%); the false
negative risk at each stage is quantified in Section~\ref{subsec:human-eval-design}.

\noindent\textbf{Threat category distribution.}
Among the 8,288 reports, Vulnerability was the most prevalent category at 50.6\% (4,194), followed by Phishing/Fraud (21.9\%, 1,817), APT/Ransomware (17.1\%, 1,420), and Malware/C2 (10.3\%, 857), as summarized in Table~\ref{tab:report-analysis}.
The dominance of Vulnerability likely reflects the high volume of CVE disclosure activity and exploit discussions on platforms such as X and Discord during the evaluation period.

\begin{table}[!tbp]
\scriptsize
\centering
\caption{Pipeline overview for the 31-day deployment (March~16--April~16, 2026)}
\label{tab:dataset-overview}
{\setlength{\tabcolsep}{15pt}
\begin{tabular}{@{}lr@{}}
\toprule
\textbf{Item} & \textbf{Count} \\
\midrule
Total Posts Collected (4 Platforms) & 873,973 (100\%) \\
\quad Telegram & 579,300 (66.3\%) \\
\quad Discord & 217,249 (24.9\%) \\
\quad X & 48,299 (5.5\%) \\
\quad Reddit & 29,125 (3.3\%) \\
\midrule
Unique Clusters Formed & 184,240 \\
Actionable TI Reports Generated & 8,288 \\
\bottomrule
\end{tabular}}
\end{table}

\begin{table}[!tbp]
\scriptsize
\centering
\caption{Threat category distribution (31-day period, 8,288 reports).}
\label{tab:report-analysis}
{\setlength{\tabcolsep}{20pt}
\begin{tabular}{@{}lr@{}}
\toprule
\textbf{Threat Category} & \textbf{Count} \\
\midrule
Vulnerability & 4,194 (50.6\%) \\
Phishing/Fraud & 1,817 (21.9\%) \\
APT/Ransomware & 1,420 (17.1\%) \\
Malware/C2 & 857 (10.3\%) \\
\midrule
Total & 8,288 (100\%) \\
\bottomrule
\end{tabular}}
\end{table}

\subsection{(1) Effectiveness of Cross-Platform Monitoring}
\label{subsec:cross-platform}

We examined whether the primary information source varies by threat category and quantified the information lost under single-platform monitoring.

\noindent\textbf{Per-category platform contribution rate.}
For each threat category, we computed the fraction of reports including posts from each platform.
The primary platform varied systematically (Table~\ref{tab:platform-specialization}):
X contributed most strongly to Phishing/Fraud (65\%); Reddit accounted for the largest
share of Malware/C2 (59\%); and Vulnerability was split between X and Discord
(46\% and 45\%).
APT/Ransomware was the most broadly distributed category, with substantial
contributions from X (47\%), Reddit (40\%), Telegram (36\%), and Discord (36\%).

\noindent\textbf{Cross-platform integration rate.}
We computed the fraction of actionable threat reports from posts spanning two or more platforms.
It was highest for APT/Ransomware (56\%), followed by Malware/C2 (46\%),
Phishing/Fraud (19\%), and Vulnerability (11\%), yielding an overall
cross-platform integration rate of 24.2\% (2,003 reports).
These 2,003 reports would have been less complete under single-platform
monitoring; the extent of this platform dependency is quantified by the loss analysis below.

\noindent\textbf{Loss rate when excluding each platform.}
Note that this metric is distinct from the cross-platform integration rate above:
whereas 24.2\% measures reports \emph{enriched} by multiple sources,
the loss rate measures reports that would \emph{disappear entirely}
because a given platform is their sole source.
Excluding X eliminates 50\% of Phishing/Fraud; excluding Telegram eliminates
29\% of Phishing/Fraud; excluding Reddit eliminates 25\% of Malware/C2 and
13\% of APT/Ransomware; and excluding Discord eliminates 43\% of Vulnerability.
Each platform thus contributes unique, non-redundant signals; TIBlender's value lies in
consolidating these platform-specific signals into actionable reports.

\begin{table}[!tbp]
\scriptsize
\centering
\caption{Per-platform contribution and cross-platform integration rates by threat category
(31-day period, actionable threat reports). Platform columns show the fraction of reports
including at least one post from each platform (may sum to $>$100\%);
Cross-plt.\ shows the fraction spanning two or more platforms.}
\label{tab:platform-specialization}
{\setlength{\tabcolsep}{3.5pt}
\begin{tabular}{@{}lrrrr|r@{}}
\toprule
\textbf{Category} & \textbf{X} & \textbf{Reddit} & \textbf{Telegram} & \textbf{Discord} & \textbf{Cross-plt.} \\
\midrule
Phishing/Fraud ($n=1{,}817$) & \textbf{65\%} & 10\% & 34\% & 10\% & 19\% \\
Malware/C2 ($n=857$) & 48\% & \textbf{59\%} & 21\% & 21\% & 46\% \\
Vulnerability ($n=4{,}194$) & \textbf{46\%} & 19\% & 3\% & 45\% & 11\% \\
APT/Ransomware ($n=1{,}420$) & \textbf{47\%} & 40\% & 36\% & 36\% & 56\% \\
\bottomrule
\end{tabular}}
\end{table}

\subsection{(2) Same-Input Comparison with IoC Extraction Baselines}
\label{subsec:baseline}

We compare TIBlender's detection coverage, novelty, and IoC quality against
 existing IoC extraction systems under a controlled same-input design:
TIBlender was given exactly the same platform-specific posts as each baseline,
eliminating any cross-platform data volume advantage.
Any performance difference therefore reflects algorithmic differences rather
than differences in input data.
 Two kinds of baselines are used: platform-specific systems, one each for X, Telegram,
and Reddit, and a platform-agnostic extractor (iocsearcher) applied uniformly to all
four platforms, including Discord.

\noindent\textbf{Baselines and input conditions.}
The baselines were Tweezers~\cite{DBLP:conf/ndss/CuiKJYKLC0L25} for X,
DarkGram~\cite{DBLP:conf/uss/RoyVKN25} for Telegram, and
Reddit2CTI~\cite{DBLP:conf/datamod/JakstaiteC23} for Reddit.
All three were evaluated on Phishing and Malware/C2
 and were selected because each can be run on exactly the same input as
TIBlender. We used the public Tweezers and DarkGram implementations with
their pretrained models, re-implemented Reddit2CTI, a non-learned pipeline
with no released implementation, from its paper description, and excluded
Discord because no comparable platform-specific system exists.
 Among the IoC extraction systems discussed in Section~\ref{subsec:related_work},
we additionally evaluate iocsearcher, the platform-independent extraction
component of GoodFATR~\cite{DBLP:journals/fgcs/CaballeroGMSSV23}, on the same
source-post text provided to TIBlender across all four platforms.
Because GoodFATR's additional filtering stages are not publicly available for
reproduction, this is a component-level extraction comparison rather than an
end-to-end evaluation of GoodFATR.
TIMiner~\cite{DBLP:journals/compsec/ZhaoYLSHL20} and
IoC Stalker~\cite{DBLP:conf/acsac/MischingerPS24} are not included: neither has a
public implementation, trained model, or labeled dataset available at the time of
writing, so a faithful same-input comparison is not possible without re-implementing
design choices that the original papers leave unspecified.

\noindent\textbf{Metrics.}
 All comparisons with baselines use TIBlender's Direct IoCs (D, extracted directly from post
text), because the baselines inspect only the input posts; IoCs added by external tool
investigation (Direct+Pivot, D+P) are reported separately as enrichment at the end of this
subsection.
Evaluation metrics are Novelty (fraction of TIBlender's IoCs absent from the baseline),
Coverage (fraction of baseline IoCs captured by TIBlender), VirusTotal (VT) detection
count distribution, and the Tranco top-1M match rate for domain-type IoCs
 (Table~\ref{tab:novelty-coverage}, Table~\ref{tab:vt-distribution}).
VT detection count indicates how many security vendors flag an IoC as malicious; because
immediate VT values can fluctuate due to definition updates and re-scan delays, we retrieve
values one week after detection~\cite{DBLP:conf/uss/ZhuSYQZS020}.
Tranco match rate, the fraction of domain-type IoCs appearing in the Tranco top-1M
list~\cite{DBLP:conf/ndss/PochatGTKJ19}, is used as an auxiliary signal for legitimate
popular domains misclassified as IoCs, and is interpreted jointly with VT distributions
because legitimate services can also be abused.

\begin{table}[!tbp]
\centering
\caption{Same-input IoC extraction comparison over
31 days (Phishing/Malware).
Nov.Ph and Nov.Mal denote the percentages of
TIBlender (D) IoCs absent from the named baseline.
Cov. denotes the percentage of baseline IoCs
captured by TIBlender (D).
Dashes indicate not applicable.}
\label{tab:novelty-coverage}

\scriptsize
\setlength{\tabcolsep}{4pt}

\begin{tabular}{@{}llrrrccr@{}}
\toprule
\textbf{Platform} &
\textbf{System} &
\textbf{Total} &
\textbf{Ph} &
\textbf{Mal} &
\textbf{Nov.Ph} &
\textbf{Nov.Mal} &
\textbf{Cov.} \\
\midrule

\multirow{3}{*}{X}
 & TIBlender (D) & 866 & 788 & 78
 & -- & -- & -- \\
 & Tweezers & 422 & 392 & 30
 & 85.7\% & 92.3\% & 28.2\% \\
 & iocsearcher & 1,114 & 1,026 & 88
 & 4.9\% & 5.1\% & 73.9\% \\

\midrule

\multirow{3}{*}{Telegram}
 & TIBlender (D) & 72 & 72 & 0
 & -- & -- & -- \\
 & DarkGram & 70 & 70 & 0
 & 86.1\% & -- & 14.3\% \\
 & iocsearcher & 150 & 150 & 0
 & 5.6\% & -- & 45.3\% \\

\midrule

\multirow{3}{*}{Reddit}
 & TIBlender (D) & 140 & 63 & 77
 & -- & -- & -- \\
 & Reddit2CTI & 277 & 85 & 192
 & 93.7\% & 87.0\% & 5.1\% \\
 & iocsearcher & 161 & 74 & 87
 & 4.8\% & 5.2\% & 82.6\% \\

\midrule

\multirow{2}{*}{Discord}
 & TIBlender (D) & 51 & 29 & 22
 & -- & -- & -- \\
 & iocsearcher & 76 & 36 & 40
 & 3.4\% & 4.5\% & 64.5\% \\

\bottomrule
\end{tabular}
\end{table}

\begin{table}[!tbp]
\scriptsize
\centering
\caption{IoC quality comparison across systems (31-day period, Phishing and Malware/C2).
 Tranco = percentage of domain-type IoCs appearing in the Tranco top-1M list.
TIBlender results use Direct (D) IoCs only.}
\label{tab:vt-distribution}
 {\setlength{\tabcolsep}{5pt}
\begin{tabular}{@{}llrrrr|r@{}}
\toprule
\multirow{2}{*}{\textbf{Platform}} &
\multirow{2}{*}{\textbf{System}} &
\multicolumn{4}{c}{\textbf{VT Detection Count}} &
\multirow{2}{*}{\textbf{Tranco}} \\
\cmidrule(lr){3-6}
& & \textbf{0} & \textbf{1--5} &
\textbf{6--10} & \textbf{11+} & \\
\midrule
\multirow{3}{*}{X}
 & TIBlender (D) & 1.6\% & 35.3\% & 36.6\% & 26.4\% & 2.3\% \\
 & Tweezers & 74.9\% & 9.2\% & 5.0\% & 10.9\% & 10.2\% \\
 & iocsearcher & 25.7\% & 26.4\% & 26.8\% & 21.2\% & 15.0\% \\
\midrule
\multirow{3}{*}{Telegram}
 & TIBlender (D) & 1.4\% & 18.1\% & 20.8\% & 59.7\% & 3.9\% \\
 & DarkGram & 87.1\% & 11.4\% & 1.4\% & 0.0\% & 44.0\%$^\ddagger$ \\
 & iocsearcher & 54.7\% & 8.7\% & 8.7\% & 28.0\% & 15.5\% \\
\midrule
\multirow{3}{*}{Reddit}
 & TIBlender (D) & 1.4\% & 30.0\% & 54.3\% & 14.3\% & 1.6\% \\
 & Reddit2CTI & 94.2\% & 5.1\% & 0.4\% & 0.4\% & 17.8\% \\
 & iocsearcher & 17.4\% & 24.2\% & 46.0\% & 12.4\% & 15.5\% \\
\midrule
\multirow{2}{*}{Discord}
 & TIBlender (D) & 2.0\% & 31.4\% & 43.1\% & 23.5\% & 2.1\% \\
 & iocsearcher & 31.6\% & 23.7\% & 28.9\% & 15.8\% & 16.2\% \\
\bottomrule
\multicolumn{7}{@{}l}{\parbox{0.9\linewidth}{$^\ddagger$ DarkGram's output contains many legitimate service domains routinely abused as contact points in fraud campaigns.}}
\end{tabular}}
\end{table}

\subsubsection{Platform-Specific Baselines (X, Telegram, Reddit)}
\noindent\textbf{Coverage and novelty.}
Using Direct IoCs only, TIBlender (D) extracted 2.1$\times$ as many IoCs as
Tweezers on X, about as many as DarkGram on Telegram (72 vs. 70), and fewer
than Reddit2CTI on Reddit (140 vs. 277).
Novelty relative to the platform-specific baselines was 85.7--93.7\% for
Phishing and 87.0--92.3\% for Malware/C2.
Thus, most of TIBlender's Direct IoCs were absent from the corresponding
baseline output; Coverage was 28.2\% (Tweezers), 14.3\% (DarkGram), and
 5.1\% (Reddit2CTI).
 TIBlender and each baseline therefore extracted substantially different IoC
sets from the same posts; the quality analysis below examines these differences.

\noindent\textbf{IoC quality.}
At least 11 VT engines detected 59.7\%, 26.4\%, and 14.3\%
of TIBlender's Direct IoCs from Telegram, X, and Reddit,
versus 0.0\%, 10.9\%, and 0.4\% for DarkGram, Tweezers,
and Reddit2CTI, respectively
(Table~\ref{tab:vt-distribution}).
Only 1.4--1.6\% of TIBlender's IoCs on these platforms
had VT\,=\,0, versus 87.1\% for DarkGram and
94.2\% for Reddit2CTI.
TIBlender's Tranco match rate was 1.6--3.9\%,
compared with 10.2\% for Tweezers and 44.0\% for DarkGram.
DarkGram included popular shortening and messaging-service
domains used in fraud, while TIBlender's residual matches
were mainly URL shorteners whose maliciousness requires
resolving the target URL.
Under identical inputs, TIBlender showed higher
VT-detection rates whether it extracted more IoCs (X),
similar numbers (Telegram), or fewer (Reddit).
These results indicate more selective extraction,
although VT and Tranco are auxiliary quality signals
and do not establish whether baseline IoCs missed
by TIBlender are false positives.

\subsubsection{Platform-Agnostic Baseline Across All Four Platforms}
\noindent\textbf{Volume and coverage.}
As a pattern-based extractor, iocsearcher returned more IoCs than
TIBlender's Direct extraction on every platform, ranging from
1.15--2.08$\times$ the TIBlender counts
(Table~\ref{tab:novelty-coverage}).
TIBlender (D) covered 45.3--82.6\% of iocsearcher's IoCs.
Conversely, across the platform/category comparisons, only
3.4--5.6\% of TIBlender (D)'s IoCs were absent from iocsearcher's
output.
Manual inspection showed that the residual TIBlender-only IoCs appeared in defanged forms that iocsearcher did not extract in our evaluation.
Thus, the small residual novelty relative to iocsearcher primarily
reflects differences in handling defanged IoC formats rather than
differences in source coverage.
Overall, the two systems exhibit substantial overlap, while TIBlender
produces a smaller, more selective IoC set; the quality analysis below
characterizes this difference.

\noindent\textbf{IoC quality.}
Although iocsearcher returned a larger number of indicators, its output
contained a substantially higher fraction of IoCs without VT detection:
17.4--54.7\% across platforms, compared with 1.4--2.0\% for
TIBlender (D).
Its Tranco top-1M match rate was also consistently higher,
15.0--16.2\%, compared with 1.6--3.9\% for TIBlender (D)
(Table~\ref{tab:vt-distribution}).
These results show that TIBlender's context-aware filtering retains
a smaller set of indicators with higher VT-detection rates and fewer
matches to popular legitimate domains.
A representative case is the Reddit reference domain
\texttt{thecybersecguru.com}, which iocsearcher extracted from a
news-style post while TIBlender excluded it after interpreting its role
in context.

\noindent\textbf{Discord.}
On Discord, where no platform-specific system exists, the same quality
difference is observed: 31.6\% of iocsearcher's IoCs had no VT detection
and 16.2\% of its domains matched the Tranco top-1M list, compared with
2.0\% and 2.1\%, respectively, for TIBlender (D).

\noindent\textbf{Pivot enrichment.}
Beyond the Direct IoCs compared above, TIBlender's external tool investigation and Pivot
expansion added IoCs originating from external evidence rather than the input posts,
bringing the Direct+Pivot (D+P) totals to 2,149 (X), 285 (Telegram), 451 (Reddit), and
175 (Discord), i.e., 2.5--4.0$\times$ the Direct counts.

\subsection{(3) Comparison with Public Threat Feeds}
\label{subsec:public-feed}

We quantify TIBlender's feed-scoped indicator absence and early detection
across all four threat categories (including Vulnerability and APT/Ransomware, which are outside the scope of the
baseline comparison) by matching against public threat feeds.

\noindent\textbf{Setup and metrics.}
We compare TIBlender's full Direct+Pivot output
against six independent feeds:
PhishTank~\cite{phishtank},
MalwareBazaar~\cite{malwarebazaar},
URLhaus~\cite{urlhaus},
CISA KEV~\cite{cisa_kev},
OTX Pulses~\cite{otx}, and
TweetFeed~\cite{tweet_feed}
(Table~\ref{tab:feed-summary}).
Feed-Scoped Absence (FSA) is the fraction of TIBlender
IoCs absent from each feed, restricted to the IoC types
and threat categories it covers; OTX Pulses and
TweetFeed are restricted by IoC type alone because
they do not specify target categories.
For overlapping IoCs, we report the fraction first
detected by TIBlender and the mean lead time among
those detected earlier.
Indicators are matched exactly after lowercasing
and URL normalization; CISA KEV's
\texttt{date\_added} is treated as midnight UTC
at day precision.

\begin{table}[!tbp]
\scriptsize
\centering
\caption{Feed-scoped absence and early detection among overlapping indicators
(31-day period). $\dagger$=feed with explicit target-category scope;
Denom.=feed-scoped IoC count; FSA=feed-scoped absence rate;
Ahead=fraction of overlapping IoCs detected by TIBlender first;
Lead Time=mean lead time (negative = TIBlender ahead).}
\label{tab:feed-summary}
{\setlength{\tabcolsep}{1.1pt}
\begin{tabular}{@{}llrrrrr@{}}
\toprule
\textbf{Feed} & \textbf{IoC Type} & \textbf{Denom.} & \textbf{Overlap} & \textbf{FSA} & \textbf{Ahead} & \textbf{Lead Time} \\
\midrule
PhishTank$^\dagger$ & domain+url & 4,914 & 227 & 95.4\% & 31.3\% & $-72$\,h \\
URLhaus$^\dagger$ & domain+url & 787 & 3 & 99.6\% & 33.3\% & $-377$\,h \\
MalwareBazaar$^\dagger$ & hash & 53 & 9 & 83.0\% & 88.9\% & $-24$\,h \\
CISA KEV$^\dagger$ & cve & 301 & 45 & 85.0\% & 17.8\% & $-94$\,h \\
TweetFeed & dom+ip+url+hash & 14,935 & 324 & 97.8\% & 16.0\% & $-157$\,h \\
OTX Pulses & cve+dom+ip+url+hash & 15,627 & 518 & 96.7\% & 13.7\% & $-109$\,h \\
\bottomrule
\end{tabular}}
\end{table}

\noindent\textbf{Absence from public feeds and early detection.}
Among the 1,126 overlapping IoCs/CVEs across all six feeds, 18.7\% were detected
by TIBlender first, confirming at operational scale that TIBlender can surface indicators before public-feed registration.
FSA ranged from 83.0\% to 99.6\%, reflecting a structural difference in data sources:
TIBlender monitors multilingual social media in real time, while public feeds rely on
researcher reports, dedicated crawlers, and manual curation.
Because feed absence alone does not establish maliciousness or correctness, we evaluate
this directly through the feed-stratified manual quality audit in
Section~\ref{subsec:human-eval-design}.
VT detection by 11 or more engines in 14.3--59.7\% of directly extracted IoCs
 (Section~\ref{subsec:baseline}) and the low fraction of VT\,=\,0 IoCs
(1.4--1.6\% in the original X, Telegram, and Reddit aggregation, measured one week after detection) provide
corroborating evidence consistent with genuine coverage gaps.

\noindent\textbf{Feed-specific trends.}
TIBlender's detection preceded PhishTank's listing for 31.3\% (71/227)
of overlapping entries, with a mean lead time of 72 hours.
TIBlender preceded TweetFeed's listings in 16.0\% and OTX Pulses' listings in 13.7\% of their overlapping cases, respectively, confirming that multilingual social media monitoring
provides complementary signals not captured by English-centric aggregation
feeds.
Campaign-level matching against OTX Pulses entries identified 51 paired
APT/Ransomware campaigns (matched by shared threat actor name or Pulse tag);
TIBlender was ahead in 23.5\% (12/51) of these by a mean of 7.2 days.
Results for MalwareBazaar and URLhaus (overlap: 9 and 3 cases,
respectively) should be interpreted cautiously given the small sample sizes.
This campaign-level rate (23.5\%) exceeds the IoC-level rate (13.7\%), suggesting that IoC-level metrics conservatively estimate TIBlender's early-warning advantage.

\noindent\textbf{CISA KEV (Exploited CVEs).}
Of 301 Vulnerability-category CVEs in the denominator, 45 overlapped with CISA KEV
(FSA 85.0\%), of which 8 (17.8\%) were captured by TIBlender before KEV
listing (mean 94 hours, maximum 13.4 days; Appendix~\ref{appendix:kev-early},
Table~\ref{tab:kev-early}).
All eight CVEs detected ahead of KEV listing were accompanied by exploitation evidence
and threat actor attribution; PoC information was also confirmed in seven of the eight.
As a complementary analysis covering all Vulnerability-category reports,
1,357 TIBlender reports reference at least one CISA KEV-listed CVE; of these,
 95 were published before the corresponding KEV listing date,
demonstrating that TIBlender captures CVEs not as mere identifiers but as attack
intelligence with context~\cite{DBLP:conf/uss/SabottkeSD15}.
This confirms pre-KEV detection as a system-wide capability rather than
isolated cases.

\subsection{(4) Human Validation and Filtering Analysis}
\label{subsec:human-eval-design}

We manually assessed report quality and filtering-stage false negatives through two
complementary analyses conducted by two annotators, both security engineers with 8 and
10 years of professional experience, respectively.

\noindent\textbf{(A) Stratified report quality audit.}
Absence from public feeds does not by itself establish maliciousness or correctness.
We therefore conducted this audit to test whether the high feed-absence rate reflects
poor report quality or genuine coverage of threats not yet registered in public feeds.
We partitioned the 8,288 generated reports into two groups: \emph{feed-overlap}
(report contains at least one IoC/CVE matched by the six public feeds evaluated in
Section~\ref{subsec:public-feed}) and \emph{feed-absent} (no IoC/CVE matched any
of those feeds under the same normalization procedure).
We then sampled 500 reports from each group for manual evaluation.
 Each annotator received every sampled report, its source posts and external-tool
results, and the metric definitions below. The annotators independently evaluated
three binary report-level metrics.
\textit{Report-level FP}: the report does not describe an actionable threat.
\textit{Unsupported IoC issue}: the report contains at least one IoC not corroborated
by any source post or external tool result (e.g., source-only domains, benign
infrastructure, or unconfirmed pivot indicators); this is a binary report-level flag,
not a ratio of unsupported IoCs among all indicators in the report.
\textit{Major quality issue}: an error in a specific report field that affects analyst
trust or operational use (e.g., unjustified severity or unsupported attribution).
Table~\ref{tab:quality-audit} shows the results
(inter-annotator agreement: 98.7\%, computed over 3,000 binary labels:
1,000 sampled reports $\times$ three metrics).
 All reported agreement rates were computed before adjudication;
disagreements in both analyses were resolved through discussion.
 Across all three metrics, the feed-absent group showed rates comparable to the
feed-overlap group: unsupported-IoC issue rates were 4.0\% vs.\ 3.6\%,
report-level FP rates were 1.8\% vs.\ 2.2\%, and major quality issues were
rare in both groups (1.4\% vs.\ 1.6\%).
The most common major quality issue was severity over-assignment in
single-source reports (9 of 15 cases).
These results suggest that the high feed-absence rate is not primarily explained
by unsupported IoCs or report-level false positives.

\begin{table}[!tbp]
\scriptsize
\centering
\caption{Manual quality audit results by group (500 reports per group, 1{,}000 total).}
\label{tab:quality-audit}
{\setlength{\tabcolsep}{1.5pt}
\begin{tabular}{@{}lrrr@{}}
\toprule
\textbf{Group} & \textbf{Report-level FP} & \textbf{Unsupported IoC} & \textbf{Major quality issue} \\
\midrule
Feed-overlap ($n=500$) & 2.2\% & 3.6\% & 1.6\% \\
Feed-absent ($n=500$) & 1.8\% & 4.0\% & 1.4\% \\
\bottomrule
\end{tabular}}
\end{table}

The 20 report-level FPs (11 feed-overlap, 9 feed-absent) were concentrated in
interpretive boundary cases, not random errors: 12 involved security-media,
vendor, or reference hosts retained as IoCs when the source post discussed an
advisory or third-party analysis rather than attacker-controlled infrastructure;
four elevated PoC or disclosure discussion to active exploitation without
sufficient corroboration; three were news-only summaries without a directly
actionable threat signal; and one was a reposted threat-intelligence item
treated as a new incident.
This suggests that the main residual errors arise from source-context
disambiguation and temporal interpretation of early discussions, rather than
from fundamental limitations in the detection pipeline.

\noindent\textbf{(B) FN evaluation at filtering stages.}
We additionally quantified false negatives introduced by the two primary filtering
stages: Step~2's severity filter and Step~4's adaptive evaluation loop.
For each stage, the same two annotators independently judged from the cluster
posts whether each excluded cluster constituted an actionable threat, using the
same actionable-threat criterion as above.
This analysis covers only these two filtering stages; we do not measure upstream losses
from collection, preprocessing, or clustering.

\emph{FN-1} (1,000 randomly sampled LOW clusters excluded by Step~2) yielded an FN
rate of 0.6\% (6/1{,}000, 95\% CI $\pm$0.5\%), inter-annotator agreement 99.3\%.
Primary causes were low-CVSS vulnerabilities later confirmed as actively exploited
(3~cases), hacktivist target list publications with minimal IoC diversity (2~cases),
and an initial access broker listing with insufficient technical indicators at
collection time (1~case).

\emph{FN-2} (1,000 randomly sampled SKIP clusters excluded by Step~4) yielded an FN
rate of 0.9\% (9/1{,}000, 95\% CI $\pm$0.6\%), inter-annotator agreement 98.6\%.
Primary causes were emerging threats with insufficient corroborating evidence
at collection time (6~cases) and single-source regional threat reports
(3~cases). These misses suggest two mitigation directions: incorporating
exploitation and threat-actor activity signals into severity scoring, and
 broadening evidence collection for emerging or region-specific threats.

\subsection{(5) Ablation Study}
\label{subsec:ablation}

We quantified the individual contribution of each component using two complementary metrics:
report count (coverage) and mean IoC/report.
All five conditions were evaluated by running TIBlender with each component disabled
in parallel with the full system over the same 31-day deployment period, enabling
direct comparison on identical input data (Table~\ref{tab:ablation}).
The five conditions are: \emph{(A1) w/o Multi-Perspective}, \emph{(A2) w/o Adaptive Eval},
\emph{(A3) w/o Trend Query}, \emph{(A4) w/o RL}, and \emph{(A5) w/o Dual-Track}
(IoC-only clustering).

\begin{table}[!tbp]
\scriptsize
\centering
\caption{Ablation results per condition (baseline: 8,288 valid reports, 2.1 IoC/report)}
\label{tab:ablation}
\begin{tabular}{@{}p{2.9cm}p{2.35cm}p{2.35cm}@{}}
\toprule
\textbf{Condition} & \textbf{Report Count ($\Delta$)} & \textbf{Mean IoC/Report ($\Delta$)} \\
\midrule
A1: w/o Multi-Perspective
 & no change
 & 2.1$\to$0.4 ($-$81\%) \\
A2: w/o Adaptive Evaluation
 & $+$3,269 \newline ($+$39.4\%, low-quality)
 & 2.1$\to$1.6 ($-$24\%) \\
A3: w/o Trend Query
 & $-$668 ($-$8.1\%)
 & no change \\
A4: w/o RL
 & no change
 & 2.1$\to$1.6 ($-$24\%) \\
A5: w/o Dual-Track
 & $-$6,398 ($-$77.2\%)
 & no change \\
\bottomrule
\end{tabular}
\end{table}

\noindent\textbf{A1: Multi-Perspective investigation.}
Disabling Multi-Perspective investigation reduces IoC yield from
2.1 to 0.4 ($-$81\%): the 1.7 IoC/report contributed by
perspective-guided investigation is eliminated, leaving only the
0.4 residual from direct IoC extraction.
Each perspective primarily discovers a distinct class of evidence (counting duplicates within each perspective before cross-perspective deduplication): Social (2.85 IoC candidates/report), 
Infrastructure (1.91), Technical (0.31), and Actor (0.31), with 
inter-perspective overlap of only 2.3\%.
The report synthesis step consolidates these candidates to 1.7 IoCs per final report through cross-perspective deduplication and quality filtering. Quality filtering removes LOW-confidence entries and indicators identified as legitimate infrastructure (e.g., CDN domains, known benign services) during context aggregation.
Report count is unaffected because cluster severity is determined
in Step~2 before perspective investigation, and Step~4 evaluates
threat relevance rather than IoC count.

\noindent\textbf{A2: Adaptive Evaluation Loop.}
Under this condition, the evaluation step is bypassed entirely: all clusters proceed directly to Step~5 without quality assessment or follow-up re-investigation.
This produces two effects, each of which reduces system-wide IoC quality.
First, the 3,269 clusters that ultimately receive a \texttt{SKIP} verdict now generate reports, averaging only 0.6~IoC/report ($+$3,269, $+$39.4\%).
Second, 845 clusters that would have undergone follow-up re-investigation and eventually proceeded with enriched context instead advance to Step~5 with only their initial investigation context, yielding lower-quality reports.
In total, 4,114 clusters are affected, and the influx of low-quality reports drives system-wide IoC yield from 2.1 down to 1.6 ($-$24\%).

\noindent\textbf{A3: Trend-based query adaptation.}
Without trend-based query adaptation, 668 reports (8.1\%) that had all their posts
sourced from trend queries disappear entirely as no supporting posts remain; a further
1,792 (21.6\%) lose their trend-query-derived posts but remain in the report count, with
reduced supporting evidence.
Because trend-query posts provide topical context rather than primary IoCs, per-report
yield is unaffected; over 31 days, the mechanism generated 15,965 dynamic
queries capturing threats beyond the reach of fixed keywords.

\noindent\textbf{A4: RL-Guided investigation.}
RL-Guided investigation delivers a quality premium: disabling it reduces system-wide
IoC yield from 2.1 to 1.6 ($-$24\%), entirely attributable to the 70.2\% of clusters
where RL applies (5,820 of 8,288).
RL-Guided mode achieved 10.0 IoC discoveries per cluster versus 6.4 without RL (+56\%). Query count increased from 19.9 to 27.1 (+36\%), yet IoC yield per query rose from 0.32 to 0.37, confirming that the gains reflect deeper investigation rather than mere query inflation.
 This comparison is against disabling RL; a static-heuristic allocation was not
evaluated as a control.

\noindent\textbf{A5: Dual-Track clustering.}
Posts about the same campaign include both IoC-bearing analyses and
context posts without IoCs.
IoC-only mode excludes all context posts (77.5\% of all posts in the 31-day deployment, measured as the fraction of posts with no extractable IoCs after Step~1 processing);
without this contextual evidence, clusters fail Step~4's quality
evaluation, losing 6,398 of 8,288 reports.
IoC yield is unaffected because context posts contain no IoCs by
definition.

These results confirm that coverage (A3, A5), IoC yield (A1, A4), and quality
control (A2) are each served by distinct components, none of which is
replaceable by another.

\subsection{Case Studies}
\label{subsec:case-studies}

We present two representative case studies: cross-platform integration 
(Case Study~1) and early warning with active exploitation context (Case Study~2).

\noindent\textbf{Case Study~1: Cross-platform integration in a CPUID supply chain
compromise.}
A HIGH-severity report titled ``CPUID Supply Chain Malware Distribution,'' generated
on April~10, 2026, integrated posts from Discord, Reddit, and X.
In isolation, Discord showed only a news article share with no clear evidence of harm;
Reddit contained fragmented victim reports of tampered CPU-Z~2.19 and HWMonitor~1.63; and
X contained scattered analysis fragments noting CDN redirection of a malicious payload.
None of these individual views revealed the full attack picture.
By integrating all three, TIBlender revealed the complete attack chain (legitimate
domain \texttt{cpuid.com} distribution links redirecting to the attacker-controlled
object-storage host \texttt{pub-\allowbreak f3252d8370f34f0d9f3b3c427d3ac33c.r2.dev}),
affected versions, and immediate response steps (stop downloads, check for infection,
block that host).
A threat that was unactionable in fragmented form became actionable only through
integration.

\noindent\textbf{Case Study~2: Early warning of API key theft via Langflow unauthenticated
RCE exploitation.}
A HIGH-severity report generated on March~23, 2026, integrated Reddit victim reports (API
key theft from compromised instances) and X alerts (attacks spreading to instances
exposing OpenAI, Anthropic, and AWS API keys).
In isolation, neither platform identified the root cause, affected vulnerability, or
remediation steps.
By integrating both, TIBlender delivered a complete picture 1.9 days before CISA KEV
listing: the \texttt{exec()} misuse enabling arbitrary code execution, exploitation within
20 hours of disclosure despite no public PoC, and three immediate response actions
(instance patching, API key rotation, and monitoring for anomalous POST requests to
public endpoints).
This converted ``knowing a vulnerability exists'' into ``active harm in progress and what
to do about it.''

\section{Discussion}
\label{sec:discussion}

\subsection{Limitations and Threats to Validity}
\label{subsec:limitations}

Our evaluation has the following limitations.

\noindent\textbf{Evaluation period and source coverage.}
The evaluation covers a 31-day window on four platforms, and results may vary across
different periods or source compositions.
Underground forums, paste sites, and less-resourced language communities remain outside
the present scope despite TIBlender's coverage of nine languages.

\noindent\textbf{Model dependency and operational cost.}
TIBlender depends on external LLM providers, and provider-side model updates could alter
the behavior of LLM-driven pipeline stages.
Using independent judges from different providers reduces dependence on a single model,
but our deployment does not measure longitudinal robustness to model-version changes.
LLM API costs are approximately \$15.6 per day, and the measured operating cost including the virtual machine and X API is approximately \$37 per day (Appendix~\ref{appendix:cost}).

\noindent\textbf{Adversarial robustness and distribution shift.}
Adversaries aware of TIBlender's monitoring may attempt evasion through decoy information
or migration to private channels; organized adversarial campaigns were not systematically
evaluated.
Gradual changes in attacker behavior are addressed by re-training the RL policy every cycle
on the preceding seven days of trajectories (BC followed by DQN fine-tuning, warm-started
from the previous policy); each update takes about six minutes on the CPU host and requires no
additional LLM calls.
Abrupt or novel threats do not rely solely on the learned policy: clusters without a
sufficiently similar historical case use Multi-Perspective mode, and RL outputs are provided
as strategy hints that agents may override.
Thus, a stale policy primarily affects investigation efficiency and IoC yield, while all
reports still pass the Step~4 evidence-quality evaluation.
The present evaluation does not establish robustness to deliberate poisoning of training
trajectories, nor does it compare the learned policy with a carefully designed static
heuristic.

\noindent\textbf{Indicator granularity on shared services.}
For resources hosted on shared services, the actionable indicator is the specific host or
object path rather than the service's parent domain; using \texttt{r2.dev} in Case Study~1
would therefore be overly broad.
Reviewing reports involving shared hosting or object-storage services, we found that most
retain the specific host or path, while a small number generalize to a parent domain.
We treat this over-generalization as a known failure mode and recommend enforcing host- or
path-level specificity in production.

\subsection{Ethical Considerations}
\label{subsec:ethics}

\noindent\textbf{Access methods.}
All four platforms were accessed through programmatic interfaces rather
than by scraping their user interfaces: the official X API v2, Reddit Data
API (OAuth via PRAW), Telegram client API (Telethon) with a registered
account, and Discord REST API. HTML was retrieved only from public
web-search results and server-directory pages used to discover Discord
invitation links.
Collection was passive and read-only: the system did not post, reply to,
or message users, respected API rate limits, and did not circumvent
CAPTCHAs or other access controls.
Posts were collected only from publicly accessible or semi-public communities;
private messages and private servers were excluded, private Telegram invitation
links were recorded but never followed, and Discord servers were screened
according to predefined rules and joined manually by an analyst.

\noindent\textbf{Terms of service.}
Collection on X, Reddit, and Telegram was conducted through
platform-provided programmatic interfaces in accordance with their
applicable terms of service.
For Discord, messages were retrieved in
read-only mode from servers manually joined by an analyst, without
bypassing access controls or interacting with users. Because Discord's
terms place restrictions on automated use of user accounts, we treat
this access pattern as a deployment limitation.

\noindent\textbf{Data handling.}
Collected posts are stored on access-controlled servers and used only for this research;
usernames and other personally identifiable information are not included in reports,
and we do not infer or report the real-world identities of platform users.
Threat-actor labels are included only when supported by source evidence.
The released source posts retain only text, platform, and timestamp, with author
identifiers and other identifying metadata removed.

\section{Conclusion}
\label{sec:conclusion}

TIBlender integrates fragmented, multilingual social media signals via a continuously adapting LLM-agent pipeline to autonomously generate actionable TI reports; it produced 8,288 reports over 31 days across four platforms.
Feed-scoped matching confirmed that 83.0\%--99.6\% of IoCs were absent
from each evaluated feed, reflecting the complementary coverage provided by
real-time social media monitoring.
Among IoCs overlapping with public feeds, 18.7\% were detected by 
TIBlender first (mean lead time: 72 hours ahead of PhishTank listings, 94 hours
ahead of CISA KEV listings).
Under identical single-platform input conditions, over 80\% of
 TIBlender's Direct IoCs were absent from each corresponding
platform-specific extractor's output, indicating that the systems
extract substantially different IoC sets from the same source posts.
Excluding any single platform eliminated up to 50\% of reports in specific threat 
categories, confirming that each platform contributes intelligence unavailable from 
the others.
The ablation study further showed that coverage, IoC yield, and quality control each 
depend on distinct components, none of which can be substituted by another.
These results establish cross-platform social media monitoring as an effective and
scalable early-warning layer for operational TI pipelines.

{\footnotesize \bibliographystyle{unsrt}
\bibliography{bibtex}}

\begin{thebibliography}{10}

\bibitem{DBLP:conf/ccs/BouwmanEHGE25}
Xander Bouwman et~al.
\newblock Can {IOCs} impose cost? the effects of publishing threat intelligence on adversary behavior.
\newblock In {\em ACM CCS}, 2025.

\bibitem{galloway2026actively}
Tillson Galloway et~al.
\newblock Actively understanding the dynamics and risks of the threat intelligence ecosystem.
\newblock In {\em NDSS}, 2026.

\bibitem{DBLP:conf/www/ShinSKLKH21}
Hyejin Shin et~al.
\newblock {\#}twiti: Social listening for threat intelligence.
\newblock In {\em WWW}, 2021.

\bibitem{DBLP:conf/raid/PaladiniFPZC24}
Tommaso Paladini et~al.
\newblock You might have known it earlier: Analyzing the role of underground forums in threat intelligence.
\newblock In {\em RAID}, 2024.

\bibitem{DBLP:conf/ndss/CuiKJYKLC0L25}
Jian Cui et~al.
\newblock Tweezers: {A} framework for security event detection via event attribution-centric tweet embedding.
\newblock In {\em NDSS}, 2025.

\bibitem{DBLP:conf/bigdataconf/NiakanlahijiSHC19}
Amirreza Niakanlahiji et~al.
\newblock {IoCMiner}: Automatic extraction of indicators of compromise from {Twitter}.
\newblock In {\em {IEEE} BigData}, 2019.

\bibitem{DBLP:journals/corr/abs-2508-10677}
Amine Tellache et~al.
\newblock Advancing autonomous incident response: Leveraging llms and cyber threat intelligence.
\newblock {\em CoRR}, abs/2508.10677, 2025.

\bibitem{DBLP:conf/uss/BuchelPLCZBE0GC25}
Marvin B{\"{u}}chel et~al.
\newblock Sok: Automated {TTP} extraction from {CTI} reports - are we there yet?
\newblock In {\em {USENIX} Security}, 2025.

\bibitem{DBLP:conf/datamod/JakstaiteC23}
Dainora Jakstaite and Ricardo~M. Czekster.
\newblock Extracting cyber threat intelligence from social media with case studies in {Twitter/X} and {Reddit}.
\newblock In {\em DataMod}, 2023.

\bibitem{DBLP:conf/IEEEares/MezziMT25}
Emanuele Mezzi et~al.
\newblock Large language models are unreliable for cyber threat intelligence.
\newblock In {\em ARES}, 2025.

\bibitem{DBLP:journals/compsec/TounsiR18}
Wiem Tounsi and Helmi Rais.
\newblock A survey on technical threat intelligence in the age of sophisticated cyber attacks.
\newblock {\em Comput. Secur.}, 2018.

\bibitem{DBLP:conf/raid/AlamBPR23}
Md~Tanvirul Alam et~al.
\newblock Looking beyond iocs: Automatically extracting attack patterns from external {CTI}.
\newblock In {\em RAID}, 2023.

\bibitem{DBLP:journals/sensors/GranadilloZD21}
Gustavo {Gonzalez Granadillo} et~al.
\newblock Security information and event management {(SIEM):} analysis, trends, and usage in critical infrastructures.
\newblock {\em Sensors}, 2021.

\bibitem{cisa_kev}
{CISA}.
\newblock {The KEV Catalog | CISA}, 2026.
\newblock \url{https://www.cisa.gov/resources-tools/resources/kev-catalog}.

\bibitem{DBLP:conf/raid/0017YL0Z20}
Jun Zhao et~al.
\newblock Cyber threat intelligence modeling based on heterogeneous graph convolutional network.
\newblock In {\em RAID}, 2020.

\bibitem{DBLP:conf/acl/JinJCCLS23}
Youngjin Jin et~al.
\newblock {DarkBERT}: {A} language model for the dark side of the {Internet}.
\newblock In {\em ACL}, 2023.

\bibitem{DBLP:journals/fgcs/CaballeroGMSSV23}
Juan Caballero et~al.
\newblock The rise of {GoodFATR}: {A} novel accuracy comparison methodology for indicator extraction tools.
\newblock {\em Future Gener. Comput. Syst.}, 2023.

\bibitem{DBLP:conf/IEEEares/Nakano0KFYHYM23}
Hiroki Nakano et~al.
\newblock Canary in {Twitter} mine: Collecting phishing reports from experts and non-experts.
\newblock In {\em ARES}, 2023.

\bibitem{DBLP:conf/uss/SabottkeSD15}
Carl Sabottke et~al.
\newblock Vulnerability disclosure in the age of social media: Exploiting {Twitter} for predicting real-world exploits.
\newblock In {\em {USENIX} Security}, 2015.

\bibitem{DBLP:conf/esorics/AlvesAGFB20}
Fernando Alves et~al.
\newblock Follow the blue bird: {A} study on threat data published on {Twitter}.
\newblock In {\em ESORICS}, 2020.

\bibitem{DBLP:conf/webi/Horawalavithana19}
Sameera Horawalavithana et~al.
\newblock Mentions of security vulnerabilities on {Reddit}, {Twitter} and {GitHub}.
\newblock In {\em WI}, 2019.

\bibitem{DBLP:conf/www/KimZKH26}
Dohee Kim et~al.
\newblock Unveiling the underground phishing ecosystem: {A} 12-year longitudinal study of deep and dark web forums.
\newblock In {\em WWW}, 2026.

\bibitem{DBLP:conf/esorics/LiZCL22}
Zhenyuan Li et~al.
\newblock Attackg: Constructing technique knowledge graph from cyber threat intelligence reports.
\newblock In {\em ESORICS}, 2022.

\bibitem{DBLP:conf/ndss/LiL24}
Zhengyi Li and Xiaojing Liao.
\newblock Understanding and analyzing appraisal systems in the underground marketplaces.
\newblock In {\em NDSS}, 2024.

\bibitem{DBLP:conf/ccs/QinXL23}
Yue Qin et~al.
\newblock Vulnerability intelligence alignment via masked graph attention networks.
\newblock In {\em ACM CCS}, 2023.

\bibitem{DBLP:conf/ccs/Bouma-SimsLC25}
Elijah~Robert Bouma{-}Sims et~al.
\newblock {'Is this a scam?'}: The nature and quality of {Reddit} discussion about scams.
\newblock In {\em ACM CCS}, 2025.

\bibitem{DBLP:conf/uss/RoyVKN25}
Sayak~Saha Roy et~al.
\newblock {DarkGram}: {A} large-scale analysis of cybercriminal activity channels on {Telegram}.
\newblock In {\em {USENIX} Security}, 2025.

\bibitem{intel471_discord_2023}
{Intel 471}.
\newblock {How Discord is Abused for Cybercrime | Intel 471}, 2026.
\newblock \url{https://www.intel471.com/blog/how-discord-is-abused-for-cybercrime}.

\bibitem{liao2016acing}
Xiaojing Liao et~al.
\newblock Acing the {IOC} game: Toward automatic discovery and analysis of open-source cyber threat intelligence.
\newblock In {\em ACM CCS}, 2016.

\bibitem{DBLP:journals/compsec/ZhaoYLSHL20}
Jun Zhao et~al.
\newblock {TIMiner}: Automatically extracting and analyzing categorized cyber threat intelligence from social data.
\newblock {\em Computers \& Security}, 95:101867, 2020.

\bibitem{DBLP:conf/acsac/MischingerPS24}
Mariella Mischinger et~al.
\newblock {{IoC} Stalker}: Early detection of indicators of compromise.
\newblock In {\em ACSAC}, 2024.

\bibitem{DBLP:conf/icde/GaoSCXLLZ21}
Peng Gao et~al.
\newblock Enabling efficient cyber threat hunting with cyber threat intelligence.
\newblock In {\em ICDE}, 2021.

\bibitem{DBLP:journals/compsec/HuZHSW24}
Yuelin Hu et~al.
\newblock {LLM-TIKG:} threat intelligence knowledge graph construction utilizing large language model.
\newblock {\em Comput. Secur.}, 2024.

\bibitem{DBLP:journals/compsec/ZhangDMWXYLC25}
Yongheng Zhang et~al.
\newblock Attackg+: Boosting attack graph construction with large language models.
\newblock {\em Comput. Secur.}, 2025.

\bibitem{strom2018attck}
Blake~E. Strom et~al.
\newblock {MITRE ATT\&CK}: Design and philosophy.
\newblock Technical Report MTR170302, {MITRE} Corporation, 2018.

\bibitem{DBLP:conf/acsac/HusariAACN17}
Ghaith Husari et~al.
\newblock {TTPD}rill: Automatic and accurate extraction of threat actions from unstructured text of {CTI} sources.
\newblock In {\em ACSAC}, 2017.

\bibitem{DBLP:conf/eurosp/ZhuD18}
Ziyun Zhu and Tudor Dumitras.
\newblock Chainsmith: Automatically learning the semantics of malicious campaigns by mining threat intelligence reports.
\newblock In {\em EuroS{\&}P}, 2018.

\bibitem{DBLP:conf/eurosp/ChengBTSG25}
Yutong Cheng et~al.
\newblock Ctinexus: Automatic cyber threat intelligence knowledge graph construction using large language models.
\newblock In {\em EuroS{\&}P}, 2025.

\bibitem{oasis2021stix}
OASIS Open.
\newblock Cyber threat intelligence technical committee, 2026.
\newblock \url{https://oasis-open.github.io/cti-documentation/}.

\bibitem{campello2015hdbscan}
Ricardo J. G.~B. Campello et~al.
\newblock Hierarchical density estimates for data clustering, visualization, and outlier detection.
\newblock {\em {ACM} Trans. Knowl. Discov. Data}, 2015.

\bibitem{wang2023selfconsistency}
Xuezhi Wang et~al.
\newblock Self-consistency improves chain of thought reasoning in language models.
\newblock In {\em ICLR}, 2023.

\bibitem{DBLP:conf/uss/ZhuSYQZS020}
Shuofei Zhu et~al.
\newblock Measuring and modeling the label dynamics of online anti-malware engines.
\newblock In {\em {USENIX} Security}, 2020.

\bibitem{DBLP:conf/ndss/PochatGTKJ19}
Victor~Le Pochat et~al.
\newblock Tranco: {A} research-oriented top sites ranking hardened against manipulation.
\newblock In {\em NDSS}, 2019.

\bibitem{phishtank}
{PhishTank}.
\newblock {PhishTank | Join the fight against phishing}, 2026.
\newblock \url{https://phishtank.org/}.

\bibitem{malwarebazaar}
{MalwareBazaar}.
\newblock {MalwareBazaar | Malware sample exchange}, 2026.
\newblock \url{https://bazaar.abuse.ch/}.

\bibitem{urlhaus}
{URLhaus}.
\newblock {URLhaus | Malware URL exchange}, 2026.
\newblock \url{https://urlhaus.abuse.ch/}.

\bibitem{otx}
{AlienVault OTX}.
\newblock {Open Threat Exchange}, 2026.
\newblock \url{https://otx.alienvault.com/}.

\bibitem{tweet_feed}
{TweetFeed}.
\newblock {Free Live IOC Feeds from Twitter/X for Infosec - TweetFeed}, 2026.
\newblock \url{https://tweetfeed.live/}.

\bibitem{mnih2015dqn}
Volodymyr Mnih et~al.
\newblock Human-level control through deep reinforcement learning.
\newblock {\em Nature}, 518, 2015.

\end{thebibliography}

\appendices

\section{External Tool Inventory}
\label{appendix:tools}

Table~\ref{tab:tools} lists Step~3 tools; $\dagger$ marks searches over the
 Step~1 internal post database rather than external APIs.

\begin{table}[h]
\scriptsize
\centering
\caption{Tools used in Step~3 ($\dagger$ = internal DB query)}
\label{tab:tools}
\begin{tabular}{@{}lp{3.0cm}p{2.1cm}@{}}
\toprule
\textbf{Tool} & \textbf{Purpose} & \textbf{Primary Perspective} \\
\midrule
Search Engine & General web search and article retrieval & All perspectives \\
RDAP/WHOIS & Domain registration and registrant attributes & Infrastructure \\
Passive DNS & DNS resolution history and related domain discovery & Infrastructure \\
Certificate Transparency & TLS certificate history (crt.sh) & Infrastructure \\
Web Crawl & Web page retrieval via proprietary crawler & Technical \\
NVD/CVE & Vulnerability database and CVSS information & Technical \\
GitHub & PoC and exploit code search & Technical \\
X & Threat information on social media & Social / Actor \\
Reddit & Community discussion and information spread tracking & Social / Actor \\
Discord$^\dagger$ & Additional keyword search over collected posts & Social / Actor \\
Telegram$^\dagger$ & Additional keyword search over collected posts & Social / Actor \\
\bottomrule
\end{tabular}
\end{table}

\section{Full TI Report Example}
\label{appendix:report-full}

\begin{figure}[!tbp]
\centering
\includegraphics[width=\columnwidth]{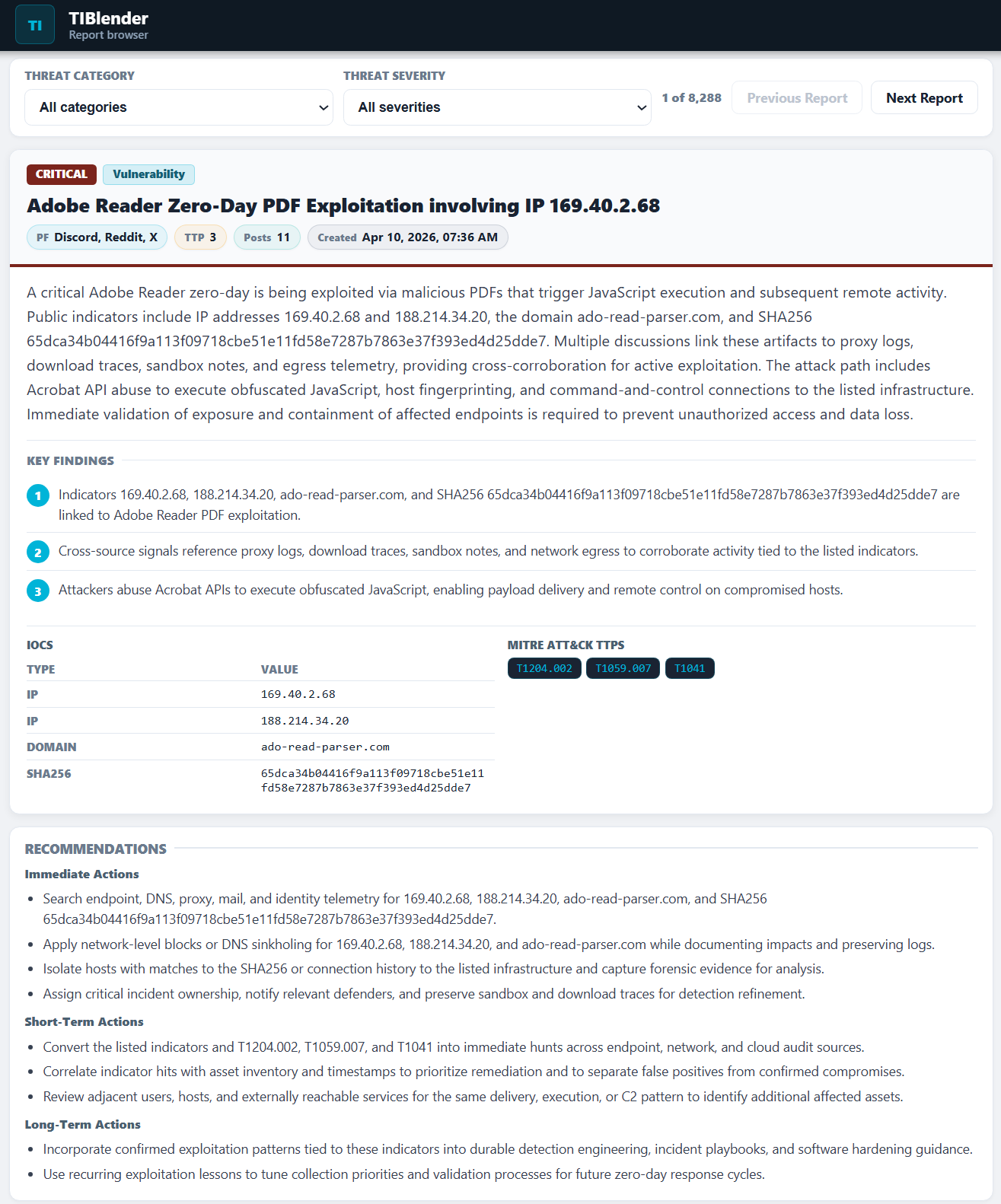}
\caption{TI report for in-the-wild exploitation of an Adobe Reader zero-day
(later assigned CVE-2026-34621), rendered in TIBlender's report viewer.
The viewer supports browsing, filtering, and searching generated reports.}
\label{fig:report-full}
\end{figure}

Figure~\ref{fig:report-full} shows a generated TI report in the browser-based viewer;
the full report and viewer are available in the public repository.

\section{Detailed Implementation Parameters}
\label{appendix:params}

\noindent\textbf{Collection keywords and supported languages (Step~1).}
Predefined keywords comprised approximately 100 terms for Phishing/Fraud, Malware/C2,
Vulnerability, and APT/Ransomware.
 English was primary, with eight additional languages: Japanese, Korean, Russian, Chinese,
Spanish, Portuguese, Arabic, and French.
 English, Russian, and Arabic were prioritized for Telegram based on pre-study analysis;
X and Reddit additionally covered Japanese, Korean, and Spanish.
Keywords combine security terminology (CVE identifier patterns, prominent malware family
names, attack technique labels, etc.) with slang, abbreviations, and hashtag formats
actually used on social media.

\noindent\textbf{IoC extraction type and confidence (Step~3).}
Collected IoCs are tagged with their extraction type and confidence level.
IoCs are classified into the Direct and Pivot extraction types defined in
Section~\ref{subsec:step3}.

Confidence (HIGH/MEDIUM/LOW) is assigned from three criteria: \emph{independent sources}
(the number of platforms or posters independently reporting the same IoC;
$\geq$3 = HIGH, 2 = MEDIUM, 1 = LOW); \emph{shared hosting} (a Passive DNS or Certificate
Transparency link to a known malicious domain raises confidence by one tier); and
 \emph{cross-tool consistency} (agreement across tools maintains confidence; contradictions
lower it by one tier).

HIGH-confidence evidence is prioritized in Step~4's sufficiency evaluation; evidence
relying solely on LOW-confidence IoCs tends to trigger \texttt{NEED\_MORE\_INFO} verdicts
in Step~4.

\noindent\textbf{Clustering (Step~2).}
HDBSCAN~\cite{campello2015hdbscan} handles unknown campaign counts,
variable-density clusters, and noise.
For LLM-normalized English summaries, we use
\texttt{BAAI/bge-large-en-v1.5}, a top-ranked open-source model on
English MTEB with 1024-dimensional embeddings.
Preliminary experiments over cosine distances 0.05--0.15 set
$\texttt{min\_cluster\_size}=3$, $\texttt{min\_samples}=2$,
and $\epsilon=0.08$ to balance over-splitting and over-merging
while retaining rare events.
An IoC Jaccard threshold $\tau_J=0.25$ links same-campaign
candidates; downstream LLM splitting resolves incorrect merges.
The non-IoC absorption threshold $\tau_{\text{abs}}=0.75$
(raised from 0.40) suppresses unrelated posts.
The severity filter retains only clusters rated MEDIUM or above
for Step~3, since LOW-only clusters did not justify investigation
cost during the 31-day deployment.

\noindent\textbf{Investigation strategy (Step~3).}
The DQN state and action space dimensions are detailed in Appendix~\ref{appendix:rl}.
The maximum simultaneous selection $K=4$ allows all four perspectives to activate in a single step.
RL-Guided mode selection uses a single similarity threshold: the similarity to the most
similar past report must be at least $\theta_{\text{rl}}=0.60$, ensuring the DQN has
learned patterns applicable to the current cluster.
 The similarity score is a logistic regression over seven features (cosine similarity,
IoC/CVE/brand Jaccard overlaps, threat category match, platform match, and temporal decay),
trained on IoC- and CVE-based objective labels.
The quality criterion for behavior cloning (BC) learning (severity $\geq$ MEDIUM) is set
as a lower bound to prevent low-quality trajectories from contaminating BC training.

\noindent\textbf{Adaptive evaluation loop (Step~4).}
The maximum number of iterations $N_{\max}=3$ was determined from the relationship between
the per-cluster processing time limit (600 seconds) and the expected benefit of additional
investigation; clusters are processed in 32 parallel workers, making the effective
throughput consistent with the one-hour cycle interval.
Preliminary experiments showed diminishing accuracy improvements beyond $N_{\max}=3$, at which point gains fell below 1\%.

\noindent\textbf{Trend detection and query adaptation (cross-cycle feedback).}
All-MiniLM-L6-v2 (384 dimensions) was chosen for speed over bge-large-en-v1.5 when
comparing large numbers of clusters each cycle.
$W=168$~h (7~days) captures weekly trends; $\tau_{\text{sim}}=0.80$ prevents spurious
historical comparisons; $k=1.5$ (approximately 6.7\% exceedance
probability under a normal approximation) balances sensitivity
and specificity; $\theta_{\text{ew}}=0.75$ targets the top 25\% by strategic value.
The one-hour cycle interval balances four-platform collection cost against real-time
timeliness; dynamic queries are passed to Step~1 for the next cycle.

\section{Implementation Details of RL-Guided Investigation}
\label{appendix:rl}

The RL model is implemented as a MultiAction DQN, which extends Deep Q-Network
(DQN)~\cite{mnih2015dqn} to support simultaneous selection of multiple actions.
The state $s_t \in \mathbb{R}^{320}$ concatenates the cluster embedding
$\phi_{\text{cluster}}$ (128-d), ongoing investigation result $\phi_{\text{context}}$ (128-d),
and investigation history $\phi_{\text{history}}$ (64-d).
The action space $\mathcal{A}$ consists of 19 action primitives (18 investigation focuses
across four perspectives plus STOP, $|\mathcal{A}|=19$).
At each step, the agent first selects the number $K$ of perspectives to activate
($1 \leq K \leq 4$) and the top-$K$ perspectives, then assigns the highest-scoring focus
from each perspective's set (4--5 focuses each), executing a composite action comprising
$K$ (perspective, focus) pairs.

\noindent\textbf{Network architecture.}
The shared encoder is a two-layer fully connected network (320$\to$256$\to$256) with
LayerNorm, ReLU, and Dropout ($p=0.1$) applied to each layer, initialized with Xavier
uniform initialization.
Four dedicated heads receive its output and independently produce: the number of actions to
select, perspective selection, attention weights over investigation focuses, and stop
decision.
Exploration uses $\epsilon$-greedy ($\epsilon_{\text{init}}=1.0$, $\epsilon_{\min}=0.01$,
decay $0.9$ per episode).
One episode corresponds to the full investigation of one cluster, with a maximum of
$T_{\max}=10$ steps per episode.

The step reward $r_t$ is designed to balance investigation depth against redundancy.
\begin{equation}
r_t = \alpha \cdot q_T \cdot \mathbf{1}[\text{terminal}]
+ \gamma_{\text{info}} \cdot g_t
- \beta \cdot c_t
- \delta \cdot p_{\text{red}}.
\label{eq:reward}
\end{equation}
Here, $q_T$ is the report quality score (range 0--1) scaled by
$\alpha=100$; $g_t = 2.0\cdot\Delta\text{IoC}+5.0\cdot\Delta\text{CVE}+3.0\cdot\Delta\text{TTP}$
is the weighted information gain per step, where $\Delta$ denotes the number of new
discoveries at that step, and $\gamma_{\text{info}}=1.0$; $c_t$ is the investigation cost
($\beta=0$; negligible at operational scale); and $p_{\text{red}}$ is the number of
duplicate actions, with $\delta=1.0$
(so the penalty term $-\delta \cdot p_{\text{red}}$ subtracts one point per duplicate action).
At the terminal step, $r_T$ is further augmented by three bonuses (not in Eq.~\ref{eq:reward}):
a coverage bonus ($10\times n_p/4$, $n_p$ = distinct perspectives activated),
an efficiency bonus ($+20$ if $<$10 actions and IoC$\geq$15;
$+10$ if $<$15 actions and IoC$\geq$10; mutually exclusive, higher tier takes priority),
and a perspective synergy bonus (sum of pair coefficients, e.g.\ Technical--Actor: 0.9).
The discount factor is $\gamma=0.99$.

\noindent\textbf{Online learning procedure.}
Before the evaluation period, the policy undergoes one week of behavior cloning (BC)
pre-training using Multi-Perspective investigation trajectories, followed by initial
DQN fine-tuning.
During the evaluation period, investigation trajectories from reports meeting the quality
criterion (severity $\geq$ MEDIUM) in the past seven days are collected at the end of each
cycle, and the policy is updated in two stages: (1) BC pre-training (20 epochs, batch size
32, learning rate $3\times10^{-4}$, early stopping patience$=3$, validation split 10\%) to
imitate high-quality trajectories, and (2) DQN fine-tuning (50 episodes, learning rate
$1\times10^{-4}$) for incremental adaptation.
Collected trajectories are stored in an experience replay buffer (capacity: 10,000
transitions); batch learning begins once 1,000 or more transitions have accumulated.
DQN fine-tuning incrementally updates from the BC-pretrained policy, and the replay buffer
retaining trajectories from multiple cycles prevents overfitting to any single cycle.

\begin{figure}[h]
\centering
\includegraphics[width=\columnwidth]{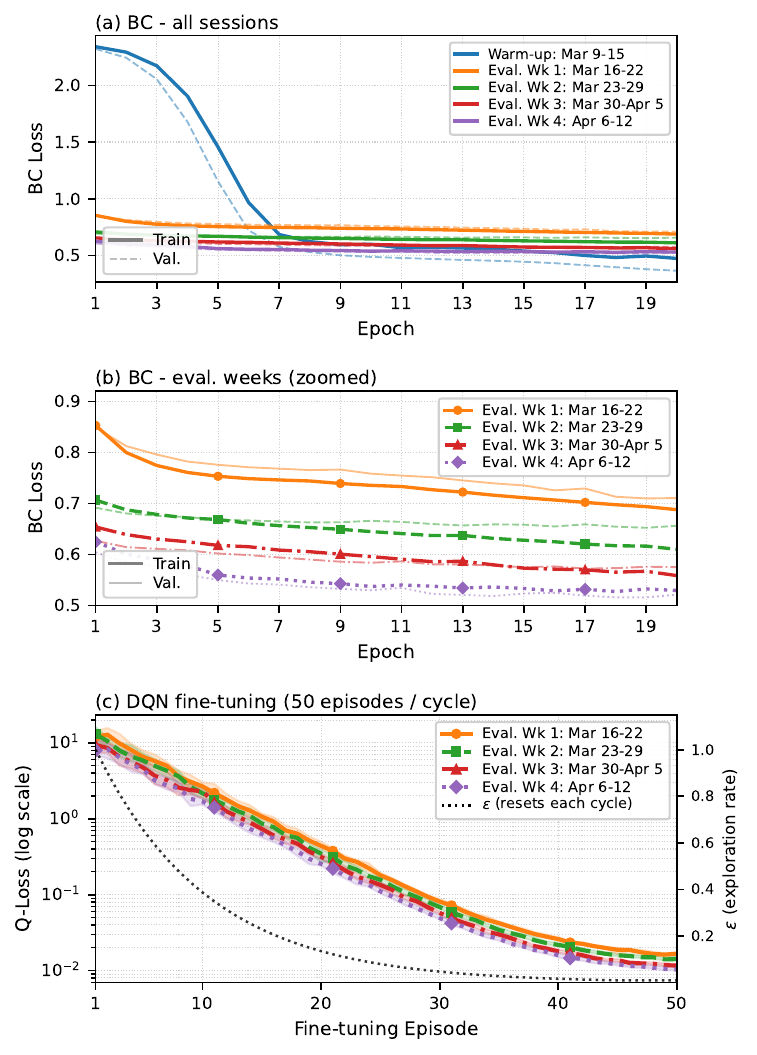}
\caption{BC/DQN convergence (March~9--April~12).
\textbf{(a)}~BC training and validation losses across
warm-up and four weekly runs (solid/semi-transparent).
\textbf{(b)}~Zoomed BC losses with distinct markers and line styles.
\textbf{(c)}~Weekly DQN Q-loss (log scale, shaded IQR)
and cycle-reset $\varepsilon$ decay (dotted, right axis).}
 
\label{fig:rl-bc-convergence}
\end{figure}

\noindent\textbf{Convergence analysis.}
Figure~\ref{fig:rl-bc-convergence} shows BC/DQN convergence.
During warm-up (March~9--15; fresh initialization, 101 trajectories),
BC loss fell from 2.34 to 0.47 in 20 epochs.
Four representative weekly runs (March~16--April~12) inherited
prior-week weights, reducing initial BC loss
(0.85 $\to$ 0.71 $\to$ 0.65 $\to$ 0.62) and its floor
(0.69 $\to$ 0.61 $\to$ 0.56 $\to$ 0.53).
Weekly replay-buffer additions were 630, 683, 666, and
$\approx$630 trajectories; validation loss closely followed training loss.
With BC initialization, DQN Q-loss dropped rapidly in the first
10 of 50 fine-tuning episodes before stabilizing.
Across weeks 1--4, initial Q-loss fell
(11.8 $\to$ 9.7 $\to$ 8.3 $\to$ 7.1), as did its floor
(0.014 $\to$ 0.012 $\to$ 0.010 $\to$ 0.009).
Exploration resets to $\varepsilon=1.0$ each cycle and decays
by 0.9 per episode toward $\varepsilon_{\min}=0.01$.

\section{LLM Agent Prompts and Output Schemas}
\label{appendix:prompts}

We present simplified system prompts (in English) and output schemas for each step's LLM
agents.
All agents prioritize Pydantic structured output; for models that do not support it, JSON
regex parsing is used as a fallback.
 \tcbset{
 agentbox/.style={
 colback=gray!4, colframe=black!50,
 fonttitle=\bfseries\scriptsize,
 left=4pt, right=4pt, top=0.2pt, bottom=0.2pt,
 breakable, boxrule=0.4pt, parbox=false
 }
}

\noindent\textbf{P1: Post Normalization Agent (Step~1).}

\begin{tcolorbox}[agentbox, title={System Prompt (simplified)}]
\scriptsize
\textit{Role:} Normalize multilingual cybercrime posts into professional English for accurate clustering.

\textbf{Critical: Verbatim IoC Preservation.}
Technical indicators MUST appear character-for-character in output.
Never paraphrase: domain \texttt{evil[.]com} stays \texttt{evil[.]com}; writing ``a malicious domain'' loses the indicator.
Applies to: domains, IPs, file hashes (MD5/SHA256), CVE IDs (\texttt{CVE-YYYY-NNNNN}), URLs, malware names.

\textbf{Required output structure (3--6 sentences):}
(1)~Attack type, (2)~Target organization/brand, (3)~Attack vector/delivery method,
(4)~Technical details (CVEs, malware family, C2 servers), (5)~Threat actor attribution,
(6)~Geographic/industry scope.
Append final line: \texttt{IoCs: <space-separated verbatim indicators>}
Non-threat post: write \texttt{Non-threat post: [brief description]}.

\textbf{Output schema:} \texttt{\{"posts": [\{"id": str, "normalized\_content": str\}]\}}
\end{tcolorbox}

\noindent\textbf{P2: Threat Clustering Agent (Step~2).}

\begin{tcolorbox}[agentbox, title={System Prompt (simplified)}]
\scriptsize
\textit{Role:} Split and refine pre-formed threat post clusters into highly specific attack campaign units.

\textbf{EXCLUDE (pre-filter):} Career/job advice, general tech discussion, fiction/creative writing,
financial market news without IoCs, security research without active incident evidence.

\textbf{INCLUDE:} Posts describing a REAL, ACTIVE security event --- ongoing attack with IoCs,
CVE with in-the-wild exploitation evidence, confirmed malicious infrastructure, fraud with victim evidence.

\textbf{Core rules:} One cluster = one concrete campaign (single target + method + infrastructure).
Target size: 3--30 posts; split if $>30$ by target org / infrastructure provider / attack sub-type.

\textbf{Cross-platform merge --- requires EXACT match on at least one of:}
same CVE ID, same threat actor name, same malicious domain/URL, same malware family.
Keyword overlap (e.g., ``Google'', ``malware'') alone is NOT sufficient.

\textbf{Severity rubric:}
CRITICAL = confirmed active exploitation + wide-scale/critical-infrastructure impact + multi-source corroboration (all three required);
HIGH = active campaign + CONFIRMED IoCs + real victim evidence;
MEDIUM = credible threat with partial evidence (default when in doubt);
LOW = unverified speculation, single unconfirmed claim, or no actionable threat evidence.

\textbf{Output schema:}
\texttt{\{"clusters": [\{"topic\_name": str, "security\_severity": "CRITICAL"|"HIGH"|"MEDIUM"|"LOW",}\\
\texttt{"topic\_description": str, "post\_ids": [str]\}]\}}
\end{tcolorbox}

\noindent\textbf{P3: Investigation Strategy Planner Agent (Step~3).}
Called once per cluster, this planner coordinates query generation
for the four perspective agents (P4--P7).

\begin{tcolorbox}[agentbox, title={System + Human Prompt (simplified)}]
\scriptsize
\textbf{System:}
``You design investigation tool queries for a threat cluster
from four perspectives: Infrastructure, Technical, Social, Actor.''

\textbf{Human (key instructions):}
Given cluster brief, focus hints, and sample posts (JSON),
design \textbf{5--15 tool queries} spanning all four perspectives.
Prioritize infrastructure investigation for any URLs found.

\textit{Query specificity constraints:}
Include only concise keywords, domains, account names, or indicators taken directly from posts.
Never output generic instructional phrases as a query.
X\_SEARCH queries MUST contain at least TWO meaningful words tied to the cluster
(brand, organization, CVE, actor, etc.); single-word tokens (e.g., ``phishing'') are invalid.

\textit{If URLs are found in posts, create queries in this priority order:}
(1)~WEB\_CRAWL for the URL (priority 9--10),
(2)~RDAP/WHOIS\_LOOKUP for the domain (priority 9),
(3)~PASSIVE\_DNS\_LOOKUP for the domain (priority 8),
(4)~TLS\_CT\_LOOKUP for the domain (priority 8).

\textbf{Per-query fields:}
perspective (Infrastructure / Social / Technical / Actor),
tool\_type (X\_SEARCH / REDDIT\_SEARCH / SEARCH\_ENGINE / WEB\_CRAWL /
RDAP/WHOIS\_LOOKUP / PASSIVE\_DNS\_LOOKUP / TLS\_CT\_LOOKUP /
NVD\_CVE\_LOOKUP / GITHUB\_SEARCH / DISCORD\_SEARCH / TELEGRAM\_SEARCH),
parameters (JSON, tool-specific),
priority (1--10), timeout\_seconds.

\textbf{Output schema:} \texttt{\{"queries": [\{perspective, tool\_type, parameters, priority, timeout\_seconds\}]\}}
\end{tcolorbox}

\noindent\textbf{P4--P7: Perspective Investigation Agents (Step~3).}
The four perspective agents run independently. Each agent instantiates the following common
templates with perspective-specific parameters and performs two subtasks: (i)~investigation
focus selection and (ii)~tool query generation.

\begin{tcolorbox}[agentbox, title={Common Prompt Templates (parameterized)}]
\scriptsize
\textbf{(i) Focus Selection:}
``You are a \{perspective\} perspective analyst.
Analyze the threat cluster and select ALL relevant investigation focuses (typically 2--4).
Being thorough is more important than being selective ---
if a focus type might provide useful information, include it.''

\textit{Input:} cluster topic, severity, description, sample posts (up to 3).

\textbf{Output schema:} \texttt{\{"selected\_focus\_types": [\{"focus\_type": str, "reasoning": str\}]\}}

\vspace{4pt}
\textbf{(ii) Query Generation:}
``You are investigating a cybersecurity threat from the \textbf{\{perspective\}} perspective.
\textbf{Perspective Focus}: \{keywords\}.
\textbf{Task}: Generate query parameters for the selected tool targeting information related to
\{focus\_types\}. Generate 1--2 high-value query parameter sets.''

\textit{Critical constraints:}
NEVER fabricate IoCs not present in cluster posts.
ONLY use domains/IPs/URLs from the Verified IoCs list.
If no verified IoCs exist, use topic-based queries WITHOUT domain/IP parameters.
Queries MUST include specific indicators (CVE ID, malware family, threat actor, or verified IoC);
generic queries (e.g., ``Discord threat'') are rejected.
RL-Guided strategy hints from past similar investigations are appended when applicable;
verified IoCs take precedence.

\textbf{Output schema (tool-dependent):}
\texttt{\{"query": str\}} for search tools;
\texttt{\{"domain": str\}} for RDAP/WHOIS and Passive DNS;
\texttt{\{"hash": str\}} for file analysis tools.
\end{tcolorbox}

The perspective-specific keywords and investigation focuses for P4--P7 are listed in
Table~\ref{tab:perspectives}.

\noindent\textbf{P8: Evaluation / Judge Agent (Step~4).}

\begin{tcolorbox}[agentbox, title={System Prompt and Judge Rubric (simplified)}]
\scriptsize
\textit{Role:} Evaluate the current multi-perspective investigation state and decide the next pipeline action.

\textbf{Evaluation 1 --- Information Sufficiency:}
SUFFICIENT = high-quality IoCs + attribution + timeline completeness;
WEAK = actionable IoCs present but material gaps remain;
INSUFFICIENT = actionable network IoCs absent.

\textbf{Evaluation 2 --- Cross-perspective Consistency:}
CONSISTENT = Infrastructure/Technical/Social/Actor perspectives tell a coherent story;
MINOR\_ISSUES = superficial differences (reconcilable, does not block report);
CONTRADICTORY = irreconcilable conflicts block report generation.

\textbf{Evaluation 3 --- Evidence Quality:}
HIGH = multi-source confirmation + official reports + recent timestamps;
MEDIUM = partial confirmation or secondary sources only;
LOW = single source or high uncertainty.

\textbf{Evaluation 4 --- Integrated Decision (choose exactly one):}
\texttt{PROCEED} --- evidence sufficient and actionable;
\texttt{NEED\_MORE\_INFO} --- evidence insufficient or irreconcilable conflicts detected; specify follow-up by perspective;
\texttt{SKIP} --- evidence insufficient to generate actionable report after investigation.

\textbf{Follow-up planning (if NEED\_MORE\_INFO):}
Specify targeted follow-up by perspective + focus\_types + tools + rationale.

\textbf{Output schema:}\\
\texttt{\{\ "sufficiency":\ \{level, reasoning, strengths, gaps\},}\\
\texttt{\ \ "consistency":\ \{level, issues: [\{claim\_a, claim\_b, is\_reconcilable, impact\}]\},}\\
\texttt{\ \ "evidence\_quality":\ \{level, reasoning, key\_factors\},}\\
\texttt{\ \ "integrated\_decision":\ str,\ "decision\_reasoning":\ str,}\\
\texttt{\ \ "follow\_up\_actions":\ [\{perspective, focus\_types, tools, rationale\}]\ \}}
\end{tcolorbox}

\noindent\textbf{P9: Report Generation Agent (Step~5).}
Uses Step~4-approved evidence to generate a STIX~2.1-compliant report.

\begin{tcolorbox}[agentbox, title={System Prompt (simplified)}]
\scriptsize
\textit{Role:} Generate comprehensive STIX-compatible threat intelligence reports in professional English.

\textbf{Severity calibration (strict thresholds):}
CRITICAL ($\leq$5\% of reports) = confirmed active exploitation + wide-scale/critical-infrastructure impact
+ multi-source corroboration --- ALL three conditions required simultaneously;
HIGH ($\leq$15\%) = active campaign with CONFIRMED IoCs + real victim evidence;
MEDIUM = credible threat with partial evidence (default --- choose MEDIUM when in doubt);
LOW = unverified speculation, single unconfirmed claim, or already-patched vulnerability.
\textit{Over-alerting degrades analyst trust; prefer MEDIUM over unjustified HIGH.}

\textbf{IoC exclusion rules (strict):}
NEVER include as IoCs: news/media site domains, cloud/CDN infrastructure (amazonaws.com, cloudflare.com, vercel.com),
URL shorteners (t.co, bit.ly), known-legitimate services,
SNS post URLs (source provenance, not attack infrastructure).
Only include IoCs verified by source evidence or tool results; discard any IoC returning 404/NXDOMAIN or clean status.

\textbf{Threat actor field:}
ONLY the adversarial actor; NEVER vendors, victims, researchers, disclosure organizations,
malware names, or campaign names. Set to Unknown if not identified.

\textbf{Output schema (11 sections):}\\
\texttt{\{cluster\_name, severity,\ summary,}\\
\texttt{\ key\_findings: [\{heading, items\}],}\\
\texttt{\ recommendations: [\{action, rationale\}],}\\
\texttt{\ iocs: [\{type, value, confidence: "HIGH"|"MEDIUM"|"LOW",}\\
\texttt{\ \ \ \ \ \ \ \ \ \ extraction\_type: "Direct"|"Pivot"\}],}\\
\texttt{\ cves, ttps, threat\_actor,}\\
\texttt{\ timeline: [\{timestamp, event, significance\}],}\\
\texttt{\ sources\}}
\end{tcolorbox}

\section{Operational Cost Estimation}
\label{appendix:cost}

\noindent\textbf{LLM API cost.}
 The daily LLM API cost is estimated from the 31-day call logs by aggregating, for each
model $m$, its input and output token counts and applying the corresponding prices:

\begin{equation}
C_{\text{daily}} \approx \frac{1}{31} \sum_{m \in \mathcal{M}}
\left( T_m^{\text{in}} \cdot p_m^{\text{in}} + T_m^{\text{out}} \cdot p_m^{\text{out}} \right),
\label{eq:cost}
\end{equation}

 where $T_m^{\text{in}}$ and $T_m^{\text{out}}$ are the input and output token counts of
model $m$ over the deployment (in millions), and $p_m^{\text{in}}$ and $p_m^{\text{out}}$
are its prices per million tokens (USD/M).
Normalization, clustering, multi-perspective investigation, report generation, and
 cross-cycle feedback use Grok 4 Fast Non-Reasoning ($p^{\text{in}}=\$0.20$/M,
$p^{\text{out}}=\$0.50$/M).
Evaluation (Step~4) runs GPT-OSS-120B and Llama 4 Maverick in parallel on the same
input, so both models' costs are summed ($p^{\text{in}}=\$0.15$/M and \$0.25/M,
$p^{\text{out}}=\$0.60$/M and \$1.00/M, respectively); the Step~4 row of
Table~\ref{tab:cost-breakdown} reports the combined token count of the two models,
which is attributed to them in equal shares when the prices are applied.
Table~\ref{tab:cost-breakdown} lists the logged token counts per pipeline step;
applying them to Equation~(\ref{eq:cost}) yields
$C_{\text{daily}} \approx \$15.60$ (\$483.6 over the 31-day deployment).

\begin{table}[!tbp]
\scriptsize
\centering
\caption{LLM token counts per pipeline step over the 31-day deployment (millions of tokens).}
\label{tab:cost-breakdown}
{\setlength{\tabcolsep}{4pt}
\begin{tabular}{@{}lrr@{}}
\toprule
\textbf{Step} & \textbf{Input tokens (M)} & \textbf{Output tokens (M)} \\
\midrule
Step 1 (Normalization) & 226.9 & 60.3 \\
Step 2 (Clustering) & 362.9 & 28.5 \\
Step 3 (Multi-perspective investigation) & 116.7 & 11.4 \\
Step 4 (Evaluation, both models) & 381.7 & 94.0 \\
Step 5 (Report generation) & 290.0 & 156.5 \\
Cross-cycle feedback & 16.2 & 2.4 \\
\midrule
Total & 1,394.4 & 353.1 \\
\bottomrule
\end{tabular}}
\end{table}

\noindent\textbf{Compute and external API costs.}
\emph{Compute:} An Azure Standard D32as v4 instance
(Section~\ref{subsec:deployment}) cost \$0.5775/h
(three-year reserved, East US; \$1.536/h pay-as-you-go).
Over 744~h, compute cost approximately \$430 (\$14/day),
excluding storage and network egress.
\emph{X API:} At \$0.005 per returned post, 48,299 reads
cost approximately \$240 (\$8/day).
Usage-based costs increase with additional keywords/accounts,
duplicate query matches, and investigation-time X searches
(Appendix~\ref{appendix:tools}).
Other external tools used free or public endpoints.
Total operating cost was approximately \$1,155 over 31 days
(\$37/day).

\section{CVEs Detected Prior to CISA KEV Listing}
\label{appendix:kev-early}

Table~\ref{tab:kev-early} lists the eight CVEs that TIBlender captured before their CISA
KEV listing date.
All eight include confirmed exploitation evidence and threat actor attribution.
\begin{table}[H]
 \scriptsize
\centering
\caption{CVEs captured by TIBlender before CISA KEV listing (first detection $<$ KEV listed; 7 of 8 with confirmed PoC)}
\label{tab:kev-early}
{\setlength{\tabcolsep}{4pt}
\begin{tabular}{@{}lllrr@{}}
\toprule
\textbf{CVE ID} &
\textbf{\shortstack[c]{First Detection\\(UTC)}} &
\textbf{\shortstack[c]{KEV Listed\\(UTC)}} &
\textbf{Lead (d)} &
\textbf{Reports$^{\ddagger}$} \\
\midrule
CVE-2026-21643 & 2026-03-30 & 2026-04-13 & 13.4 & 24 \\
CVE-2026-34197 & 2026-04-08 & 2026-04-16 & 7.1 & 12 \\
CVE-2026-1340 & 2026-04-03 & 2026-04-08 & 4.9 & 7 \\
CVE-2026-33017 & 2026-03-23 & 2026-03-25 & 1.9 & 34 \\
CVE-2026-34621 & 2026-04-11 & 2026-04-13 & 1.4 & 6 \\
CVE-2026-33634 & 2026-03-24 & 2026-03-26 & 1.2 & 3 \\
CVE-2026-3502 & 2026-03-31 & 2026-04-02 & 1.1 & 8 \\
CVE-2026-35616 & 2026-04-05 & 2026-04-06 & 0.4 & 1 \\
\bottomrule
\end{tabular}}
 \par\smallskip\noindent{\footnotesize
 $^{\ddagger}$ TIBlender reports published before the CVE's KEV listing date.}
\end{table}

The eight CVEs generated 95 TIBlender reports before their KEV listing dates.
Across all 4,194 Vulnerability-category reports, 34.4\% contain active exploitation
evidence, 20.9\% include threat actor attribution, 9.4\% reference PoC code, and 35.2\%
are high-priority (score $\geq$ 2)~\cite{DBLP:conf/uss/SabottkeSD15}.

\end{document}